\pdfoutput=1
\documentclass[12pt]{article}
\usepackage{fullpage}
\usepackage{amsmath,amssymb,amsfonts}
\usepackage{epsfig,subfigure}
\usepackage{graphicx}

\newcommand\be{\begin{equation}}
\newcommand\ee{\end{equation}}
\newcommand\bes{\begin{eqnarray}}
\newcommand\ees{\end{eqnarray}}

\newcommand{\old}[1]{{}}

\newcommand\curl{\nabla\times}  

\newcommand\B{\mathbf{B}}
\newcommand\Bxy{\mathbf{B}^{xy}}
\newcommand\rBxy{\widetilde{\mathbf{B}}^{xy}}
\newcommand\E{\mathbf{E}}

\newcommand\vel{\mathbf{u}}

\newcommand\U{\mathbf{U}}
\newcommand\F{\mathbf{F}}
\newcommand\G{\mathbf{G}}
\newcommand\V{\mathcal{V}}
\newcommand\K{\mathcal{K}}

\def\bx{\mathbf x}

\numberwithin{equation}{section}

\title{Divergence-Free WENO Reconstruction-Based Finite Volume Scheme for Solving Ideal MHD Equations on Triangular Meshes}

\author{
Zhiliang Xu \footnotemark[3] and Dinshaw Balsara \footnotemark[4]}

\begin{document}

\maketitle

\renewcommand{\thefootnote}{\fnsymbol{footnote}}
\footnotetext[3]{({\tt E-mail: zxu2@nd.edu})
\\
Department of Applied and Computational Mathematics and Statistics,
University of Notre Dame, Notre Dame, IN
46556. Research was supported in part by NSF grant DMS-0800612.}
\footnotetext[4]{({\tt E-mail: Dinshaw.S.Balsara.1@nd.edu})
\\
Department of Physics, University of Notre Dame, Notre Dame, IN
46556. Research was supported in part by NSF grant AST-1009091.}
\renewcommand{\thefootnote}{\arabic{footnote}}

\begin{abstract}
In this paper, we introduce a high-order accurate
constrained transport type finite volume method to solve
ideal magnetohydrodynamic equations on two-dimensional triangular meshes.
A new  divergence-free WENO-based reconstruction method is developed to
maintain exactly divergence-free  evolution of the numerical magnetic field.
A new weighted flux interpolation approach is also developed to
compute the $z$-component of the electric field at vertices of grid cells.
We also present numerical examples to demonstrate the accuracy and robustness of the
proposed scheme.
\end{abstract}


\section{Introduction}

The ideal MHD equations model the dynamics of an electrically conducting fluid. Numerical solutions to magnetohydrodynamic (MHD) equations are of great importance to many applications in astrophysics and engineering.
Many efforts in solving the ideal MHD equations numerically have focused on the divergence-free evolution of the magnetic field implied by the
induction equation
\be
\frac{\partial \B}{\partial t} +  \curl \E =0 ~.
\ee
Here $\B$ is the megnetic field, and $\E$ is the electric field defined by
$\E = - \vel \times \B  $ for ideal MHD. $\vel$ is the velocity. $\mathbf{J} = \curl \B$ is the current density.  The induction equation ensures that the magnetic field remains divergence-free if it is
divergence-free initially. In numerical simulations, maintaining discrete divergence-free is also important.
Previous studies \cite{BraBar80,BalKim04} have shown that a divergence error on the order of numerical truncation error introduced by the numerical scheme can lead to spurious solutions and the production of negative pressures.

To name a few methods to ensure divergence-free evolution of the magnetic field, these include Hodge projection approach \cite{ZacMal94}, Powell's source term formulation \cite{Powell94}, locally divergence-free discontinuous Galerkin (DG) method \cite{LiShu05, CocLi04}, constrained transport (CT) methods \cite{Yee66, BreLyo81, RyuJon95, DaiWoo98, RyuMin98, BalSpi99b,BalSpi99, Balsara04, GarSto05}, generalized Lagrange Multiplier method \cite{DedKem02}, and many others \cite{BraBar80,LiXu11,Tot00}.

Despite these advances, almost all previous works have been focused on structured meshes. The CT type divergence-free formulation on structured meshes has been achieved at the second order accuracy in \cite{Balsara01,Balsara04} and higher order accuracy in \cite{Balsara09}.
Several problems with complex geometry require the use of unstructured meshes. It is, therefore, desirable to design high order accurate divergence-free formulation for unstructured meshes.

For the CT type formulation on structured meshes, the second-order accurate representation of the magnetic field at the cell center can always be obtained by averaging the facial magnetic field. However, for the unstructured meshes, this is much harder to do, as there is no concept of arithmetic averaging of facial magnetic field to the center of the grid cells. As a result, the zone averaged magnetic field has always to be obtained via a reconstruction process on unstructured meshes. This makes divergence-free MHD on unstructured meshes slightly more challenging than the same process on structured meshes.

In this paper, we introduce a divergence-free WENO reconstruction-based finite volume scheme up to the third order accuracy for solving ideal MHD equations on two-dimensional triangular meshes. ENO and WENO finite volume schemes have been introduced in many previous works for solving scalar conservation laws as well as compressible hydrodynamical flow problems using unstructured meshes \cite{HarCha91,Abl94,Son97, HuShu99, KasIsk05, KasIsk05, DumKas07}. However, to the best of our knowledge, divergence-free high order ($> 2$) accurate
finite volume schemes for solving ideal MHD equations on triangular meshes have not yet been available.
To satisfy the divergence-free constraint of the magnetic field, we employ the CT framework. The basic idea of the CT framework adopted in the present paper is to introduce a staggered magnetic
field at cell edges in two spatial dimensions (2D) (or faces in three spatial dimensions) and a staggered electric field at cell  corners   (or edges in three spatial dimensions) so that the computed magnetic field conserves a discrete definition of the divergence. To achieve this, a weighted flux interpolation approach based on \cite{BalSpi99b} is introduced in this paper to compute the $z$-component of the electric field. To achieve high order accuracy, a new divergence-free WENO reconstruction method is introduced to reconstruct a cell centered magnetic field from the staggered allocated magnetic field on cell edges in two spatial dimensions. Additionally, the reconstructed piecewise smooth magnetic field is consistent at a cell edge by having the same cell edge-length-averaged value of normal component of the magnetic field when
evaluated by using reconstructed magnetic field supported on triangles sharing this
edge respectively.
For the cell centered variables, the WENO reconstruction described in \cite{KasIsk05,DumKas07} is utilized.
Numerical experiments show that the present divergence-free WENO reconstruction-based finite volume scheme is robust and accurate.

The paper is organized as follows. Section \ref{sec:FV} describes
 the CT type finite volume formulation to solve the ideal MHD equations. We start with introducing governing equations, notations for domain partition and discretization.
 Specifically, the proposed weighted flux interpolation approach to compute the $z$-component of the electric field is described in
subsection \ref{sec:E_z}.
Section \ref{sec:WENO} describes the proposed reconstruction algorithm.
The second-order accurate and the third-order accurate divergence-free WENO reconstruction methods are catalogued in detail in subsection \ref{sec:B_reconstruct}.  Numerical tests are given in Section \ref{sec:test}
to demonstrate the accuracy and non-oscillatory properties of the
proposed scheme by computing smooth solution and shock wave related problems. We draw conclusions in Section \ref{sec:conc}.

\section{Finite Volume Formulation}
\label{sec:FV}

Ideal MHD governing equations in the conservation form can be expressed as
\be
 \partial_t \U + \partial_x \F(\U) + \partial_y \G(\U) = 0~,
\label{eq:2d_consv}
\ee
where
\be
\U = \left(\rho, \rho u_x, \rho u_y, \rho u_z, \varepsilon, B_x, B_y, B_z  \right)^T,
\ee
and
\be
\begin{array}{ll}
\F(\U) = \left(
\begin{array}{l}
\rho u_x \\
\rho u_x^2 + p - B_x^2\\
\rho u_x u_y - B_x B_y\\
\rho u_x u_z - B_x B_z\\
(\varepsilon +p)u_x - B_x(\vel\cdot \B) \\
0 \\
(u_x B_y - u_y B_x) \\
- (u_z B_x  - u_x B_z)
 \end{array}
 \right)~,
 &
\G(\U) = \left(
\begin{array}{l}
\rho u_y \\
\rho u_x u_y - B_x B_y\\
\rho  u_y^2 +p - B_y^2\\
\rho u_y u_z - B_y B_z\\
(\varepsilon +p)u_y - B_y(\vel\cdot \B) \\
-(u_x B_y - u_y B_x) \\
0 \\
(u_y B_z - u_z B_y)
 \end{array}
 \right)~.
\end{array}
\ee

Here $p = p_{gas} + \B\cdot \B/2$ is the total pressure, $p_{gas}$ is the gas pressure that satisfies
the following equation of state
$$
p_{gas} = (\gamma-1)(\varepsilon - \frac{1}{2} \rho \vel\cdot \vel - \frac{1}{2} \B \cdot \B)~,
$$
with $\vel = (u_x, u_y, u_z)^T$ and $\B = (B_x, B_y, B_z)^T$.
For a  2D ideal MHD problem, we have
\be
E_z = -u_x B_y + u_yB_x~.
\label{eq:E_z_F}
\ee

We employ the CT approach and the Godunov type finite volume scheme to solve  Eq. (\ref{eq:2d_consv}). To this end,
the physical domain $\Omega$ is partitioned into a collection
of $\mathcal{N}$ triangular cells $\K_i$ so that $\Omega = \bigcup^{\mathcal{N}}
_{i=1} \mathcal{K}_i$ and we define
\begin{equation}
\mathcal{T}_h = \{\mathcal{K}_i: i = 1, \cdots,\mathcal{N}\}~.
\label{eq:partition}
\end{equation}
We also collect cell edges $\mathcal{L}_j$ to form
\be
\mathfrak{E}_h = \{\mathcal{L}_j: j = 1, \cdots, \mathcal{N}_\mathfrak{E} \}~,
\ee
where $\mathcal{N}_\mathfrak{E}$ is the total number of edges in the partition.
For every cell edge $\mathcal{L}_j$, we uniquely identify an edge unit normal $\mathbf{n}_j$ and tangent $\boldsymbol{\zeta}_j$. Here $\boldsymbol{\zeta}_j$ is obtained by rotating $\mathbf{n}_j$ 90 degrees in the counterclockwise direction.
For simplicity, we assume that there are no hanging nodes in the partition $\mathcal{T}_h$. Let the edges of cell $\K_i$ be denoted as $\partial \K_{i, \ell}, ~\ell = 1, 2, 3.$ For convenience in discussion, we define a mapping between the local cell edge index $\ell$ of cell $\K_i$ and the global edge index $j$ such that
\be
\ell = \ell_i(j)~~~~ {\rm and }~~~~ j = \ell^{-1}_i(\ell)~.
\label{eq:edge_map}
\ee
We also define the mesh parameter $h$ to be
\be
\begin{array}{lllll}
h_{\K_i} & = & {\rm the ~diameter ~ of ~\K_i} & = & {\rm the ~ longest ~ side ~ of ~\K_i} \\
h & = & \max_{\K_i \in \mathcal{T}_h  } h_{\K_i}~ & & .
\end{array}
\label{eq:mesh_par}
\ee

We place the magnetic field variables $B_x$ and $B_y$ at the cell edges
to maintain the global
divergence-free evolution of the magnetic field;  the $z$-component of the electric field $E_z$ at the cell vertices; and the conservative variables $\rho, \rho\vel$ and $\varepsilon$ and $B_z$ on the cells. $B_x$ and $B_y$ are always initialized to be divergence-free.
The Godunov type finite volume scheme is utilized to evolve $\rho, \rho\vel$, $\varepsilon$
and $B_z$ on the cells and the normal component of the magnetic field within the $xy$-plane on the cell edges. To evaluate $E_z$ at cell vertices, the flux-interpolated approach introduced by Balsara and Spicer \cite{BalSpi99b} is further developed here.

For convenience in discussion, we introduce  notations $\U^H = (\rho, \rho\vel, \varepsilon, B_z)^T$ and $\B^{xy} = (B_x, B_y)$ so that
\be
 \partial_t \U^H + \partial_x \F^H(\U) + \partial_y \G^H(\U) = 0~,
\label{eq:2d_hydo}
\ee
where
\be
\begin{array}{ll}
\F^H(\U) = \left(
\begin{array}{l}
\rho u_x \\
\rho u_x^2 + p - B_x^2\\
\rho u_x u_y - B_x B_y\\
\rho u_x u_z - B_x B_z\\
(\varepsilon +p)u_x - B_x(\vel\cdot \B) \\
-(u_z B_x - u_x B_z)
 \end{array}
 \right)~, &
 \G^H(\U) = \left(
\begin{array}{l}
\rho u_y \\
\rho u_x u_y - B_x B_y\\
\rho  u_y^2 +p - B_y^2\\
\rho u_y u_z - B_y B_z\\
(\varepsilon +p)u_y - B_y(\vel\cdot \B) \\
u_y B_z - u_z B_y
 \end{array}
 \right) ~.
\end{array}
\ee

And
\be
 \partial_t \B^{xy} + \partial_x \F^B(\U) + \partial_y \G^B(\U) = 0~,
\label{eq:2d_mag}
\ee
where

\be
\begin{array}{ll}
\F^B(\U) = \left(
\begin{array}{l}
0 \\
u_x B_y - u_yB_x \\
 \end{array}
 \right)~, &
\G^B(\U) = \left(
\begin{array}{l}
-u_x B_y + u_yB_x \\
0 \\
 \end{array}
 \right)~.
\end{array}
\ee
Thus solving Eq. (\ref{eq:2d_consv}) is equivalent to solving
equations (\ref{eq:2d_hydo}) and (\ref{eq:2d_mag}) together.


\subsection{Semi-discrete finite volume scheme for the cell-centered $\U^H$}
\label{sec:FV_H}

Taking the cell $\mathcal{K}_i$, $i = 1, \cdots, \mathcal{N}$, in partition (\ref{eq:partition}) as a discrete control volume, the semi-discrete
finite volume method for solving Eq. (\ref{eq:2d_hydo}) is formulated by
integrating (\ref{eq:2d_hydo}) over the cell $\mathcal{K}_i$:
\be
\frac{d}{dt} \overline{\U}^H_{k,i}(t) +
\frac{1}{|\mathcal{K}_i|} \int_{\partial \mathcal{K}_i} (\F^H_k, \G^H_k)\cdot \mathbf{n}_i d\Gamma = 0~,
\label{eq:semi_hydro}
\ee
where $\overline{\U}^H_{k,i}(t)$ is the cell average of the $k^{th}$ ($k = 1, \cdots, 7$) component of
$\U^H$ on $\mathcal{K}_i$, $\F^H_k$ is the $k^{th}$ component of $\F^H$,
$\G^H_k$ is the $k^{th}$ component of $\G^H$,
 and $\mathbf{n}_i$ is the
outward unit normal of the boundary of the cell $\mathcal{K}_i$. $|\mathcal{K}_i|$ is a shorthand notation for the area of $\K_i$.

To solve Eq. (\ref{eq:semi_hydro}) numerically, we evaluate the flux integral by Gaussian quadrature rule with the exact value of $(\F^H, \G^H) \cdot \mathbf{n}_i$
being replaced by the Lax-Friedrichs flux  $\F^*(x,y,t)$ given by
\be
\F^*_k(x,y,t) = \frac{1}{2}\left[ (\F^H_k(\U^{-}),  \G^H_k(\U^{-})) + (\F^H_k(\U^{+}),  \G^H_k(\U^{+}))\right] \cdot \mathbf{n}_i - \frac{\alpha}{2}(\U_k^{H,+}-\U_k^{H,-}).
\ee
Here $\alpha$ is taken as an upper bound for the eigenvalues of the Jacobian in the $\mathbf{n}_i$ direction; 
$\U^{-}$ (or $\U^{H,-}$) and $\U^{+}$ (or $\U^{H,+}$) are the numerical values of $\U$ (or $\U^H$) inside the triangle and outside the triangle at the Gaussian point. To this end,
we obtain the following semi-discrete finite volume  scheme for solving Eq. (\ref{eq:2d_hydo})
\be
\frac{d}{dt} \overline{\U}^H_{h,k,i}(t) +
\frac{1}{|\mathcal{K}_i|} \int_{\partial \mathcal{K}_i}  \F^*_k d\Gamma = 0~,
\label{eq:hydro_scheme}
\ee
where $\overline{\U}^H_{h,k,i}(t)$ is the approximate cell average of the $k^{th}$ component of
$\U^H$  on the cell $\mathcal{K}_i$.


\subsection{Semi-discrete finite volume scheme for the edge-centered normal component of $\B^{xy}$}
\label{sec:FV_B}

The 2D constrained transport scheme developed in the present paper
is based upon cell edge-length-averaged magnetic field located at
the edges of grid cells.
On every cell edge $\mathcal{L}_j \in \mathfrak{E}_h $, we solve  Eq. (\ref{eq:2d_mag})
to evolve the normal component of $\Bxy$ with respect to the defined cell edge unit normal $\mathbf{n}_j$ .
Denote the normal and tangential contribution of
$\Bxy$ in directions given by $\mathbf{n}_j$ and $\boldsymbol{\zeta}_j$ to be
$B_n$ and $B_\zeta$ respectively.
We rewrite Eq. (\ref{eq:2d_mag}) by $B_n$ and $B_\zeta$ to obtain
\be
\partial_t\left(
\begin{array}{l}
 B_n \\
 B_\zeta
\end{array}
\right )
+
\partial_n\left(
\begin{array}{l}
 0 \\
 u_n B_\zeta - u_\zeta B_n
\end{array}
\right )
+
\partial_\zeta\left(
\begin{array}{l}
 -u_n B_\zeta + u_\zeta B_n \\
  0
\end{array}
\right )
=0~.
\label{eq:B_nor_tan}
\ee
Here $u_n$ and $u_\zeta$ are the components of velocity $\vel$ in the $\mathbf{n}_j$
and $\boldsymbol{\zeta}_j$ directions respectively.

Let $\overline{B_{n,j}}$ be the edge-length-averaged $B_n$ on the edge $\mathcal{L}_{j}$
defined by
\be
\overline{B_{n,j}} = \frac{1}{|\mathcal{L}_{j}|} \int_{\mathcal{L}_{j}} B_n d\zeta~,
\ee
where $|\mathcal{L}_{j}|$ is a shorthand notation for the length of the edge $\mathcal{L}_{j}$.
Integrating  Eq. (\ref{eq:B_nor_tan}) along the cell edge $\mathcal{L}_{j}$,  the semi-discrete finite volume scheme to evolve $\overline{B_{n,j}}$ numerically on $\mathcal{L}_{j}$
can be expressed as
\be
\frac{d}{dt} \overline{B_{h,n,j}} = -\frac{E_z(\mathcal{L}_{j,e}) - E_z(\mathcal{L}_{j,s}) }{|\mathcal{L}_{j}|}~,
\label{eq:semi_Bnor}
\ee
since
$$
E_z = -u_n B_\zeta + u_\zeta B_n ~.
$$
Here $\overline{B_{h,n,j}}$ is the approximate cell edge-length-averaged $B_n$ on $\mathcal{L}_{j}$. $E_z(\mathcal{L}_{j,e})$ is numerical approximation of the $z$-component of $\E$ at the end point of $\mathcal{L}_{j}$, and $E_z(\mathcal{L}_{j,s})$ is numerical approximation of the $z$-component of $\E$ at the starting point of $\mathcal{L}_{j}$.
In the direction of $\boldsymbol{\zeta}_j$, the two end points of the edge $\mathcal{L}_j$ are defined to be the starting and the end point of $\mathcal{L}_j$ respectively.
The method to
compute $E_z$ is described in Section \ref{sec:E_z}.


\subsection{Computing $E_z$ at the vertices of cells by flux interpolation}
\label{sec:E_z}

\begin{figure}[ht]
\begin{center}
\includegraphics[width=2.0in]{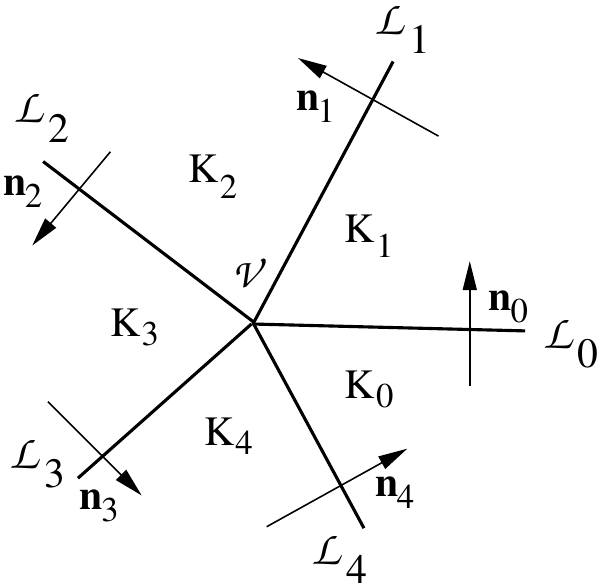}
\caption{The stencil to compute $E_z$ at the vertex $\V$. Cell edges
$\mathcal{L}_0, \cdots, \mathcal{L}_4$ which separate
triangular cells
$\K_0, \cdots, \K_4$ all have one end at $\V$.
 $\mathbf{n}_0, \cdots, \mathbf{n}_4$ are unit normals of these edges respectively.
\label{fig:Ez_stencil}
}
\end{center}
\end{figure}

In our scheme, one has to obtain the electric field $E_z$ at vertices of the triangular mesh (see Fig. \ref{fig:Ez_stencil}). In \cite{BalSpi99b}, it was shown that there is a dualism between the electric field and the properly upwinded flux. In fluid dynamics, such a flux takes on contributions that are upwinded normal to a zone face. For MHD, the electric field at the vertex $\V$ in Fig. \ref{fig:Ez_stencil} should take on properly upwinded contributions from all possible directions. This necessarily would require a multi-dimensional Riemann solver. For structured meshes, such a multi-dimensional Riemann solver has been presented in \cite{Bal10}. Unfortunately, a multi-dimensional Riemann solver that works for MHD on unstructured meshes has not been presented in the literature.  For that reason, we use the available ideas on multi-dimensional upwinding from \cite{BalSpi99b} and the idea of doubling dissipation in each direction from \cite{LonDel04, GarSto05}.

Below we describe an algorithm to compute the $z$-component of the electrical field $E_z$
at vertices of the mesh. The algorithm results in an upwinded
choice of $E_z$ in a multi-dimensional fashion.

See Fig. \ref{fig:Ez_stencil}.  Suppose triangles $\K_0, \cdots, \K_4$ meet at the vertex $\V$. The edges shared by triangles are labeled by $\mathcal{L}_0, \cdots, \mathcal{L}_4$ and the associated unit normals of edges by  $\mathbf{n}_0, \cdots, \mathbf{n}_4$ respectively.

On the each edge $\mathcal{L}_l, l = 0, \cdots, 4 = n_v$, using Eq. (\ref{eq:E_z_F}), which shows the dualism between $E_z$ and flux, we
obtain
\be
E_{z,l}(x_{\V}, y_{\V}) = - \F_{LF,6}\left(\U^{-}(x_{\V}, y_{\V}), \U^{+}(x_{\V}, y_{\V})\right)~,
\ee
from numerical flux interpolation. Here $(x_{\V}, y_{\V})$ are the coordinates of $\V$.
 $\F_{LF}$ is the Lax-Friedrichs flux with double dissipation for solving Eq. (\ref{eq:2d_consv}); and
$\F_{LF,6}$ is the $6^{th}$ component of $\F_{LF}$. Let $\K_l^{\rm int}$ denote the interior of cell $\K_l$.
$$
\U^{-}(x_{\V}, y_{\V}) = {\rm lim}_{(x, y) \rightarrow (x_{\V}, y_{\V})} \U(x,y,t)~,
~~~~~ {\rm where}~~ (x, y)\in \K_{l}^{\rm int}~,
$$
$$
\U^{+}(x_{\V}, y_{\V}) = {\rm lim}_{(x, y) \rightarrow (x_{\V}, y_{\V})} \U(x,y,t)~,
~~~~~{\rm where}~~ (x,y) \in \K_{(l+1)\%(n_v+1)}^{\rm int}~.
$$
 Thus
\be
\F_{LF}(\U^{-}, \U^{+}) = \frac{1}{2} \left[ (\F(\U^{-}), \G(\U^{-}))  +
(\F(\U^{+}), \G(\U^{+}))
\right] \cdot \mathbf{n}_{s} - \alpha (\U^+ - \U^-)~.
\ee
Here $\alpha$ is taken as an upper bound for the eigenvalues of the Jacobian in the $\mathbf{n}_s$ direction.

If the flow is locally smooth, we can  take the arithmetic average
$$
E_z(x_{\V}, y_{\V}) = \frac{1}{1+n_v}\sum^{n_v}_{l=0} E_{z,l}(x_{\V}, y_{\V})~
$$
to obtain a unique $E_z$ at the vertex $\V$. However,when discontinuities are present it is beneficial to allow  the evaluation of $E_z$ to locally adjust to those discontinuities. To achieve this, we design switches to detect strong magnetosonic shocks and strongly compressive motions and the direction of propagation of the discontinuity.

For this purpose, we first use a least square approach to construct linear profiles of pressure and
velocity at the vertex $\V$ respectively as follows. See Fig. \ref{fig:Ez_stencil}. Briefly, we first compute on $\K_l$, $l = 0, \cdots, 4$, pressure $p_{gas,l}$ and velocity $(u_{x,l}, u_{y,l})$ from cell average values of conservative variables at the cell centers. Let $\mathbf{v} = (p_{gas}, u_x, u_y)^T$; and the $s^{th}$ component ${v}_s(x,y)$, $s = 1, 2, 3$, of $\mathbf{v}$ be represented by a linear polynomial
\be
v_s(x,y) = v_{0,s} + \frac{\triangle_x v_s}{h} (x - x_{\V}) + \frac{\triangle_y v_s}{h} (y - y_{\V})~
\ee
where $h$ is the mesh parameter; $(\triangle_x v_s, \triangle_y v_s)$ is the
undivided difference approximation to the gradient of the exact profile of pressure or velocity
at $\V$.
Parameter values $v_{0,s}$,  $\triangle_x v_s$, $\triangle_y v_s$ are determined by solving
$$
\left\{
v_{0,s} + \frac{\triangle_x v_s}{h} (x_l - x_{\V}) + \frac{\triangle_y v_s}{h} (y_l - y_{\V}) = v_{s,l}~, ~~~~~~~~~~l = 0, \cdots, 4
\right .
$$
in the least square sense. Here $(x_l, y_l)$ are the coordinates of
the cell center of $\K_l$;  $v_{s,l}$ stands for the value of the $s^{th}$ component of $\mathbf{v}$ on cell $\K_l$, which are $p_{gas,l}$,  $u_{x,l}$ and $u_{y,l}$ respectively.

The first switch, \textbf{SW1}, which is used to pick out strong magnetosonic shocks or
configuration that may develop into such a shock, is accomplished by taking the undivided
gradient of the pressure at the vertex $\V$ and comparing it with the minimum pressure in the vicinity. \textbf{SW1} is switched on if
\be
|\triangle_x p_{gas}|(x_{\V}, y_{\V}) + |\triangle_y p_{gas}|(x_{\V}, y_{\V}) > \beta \min (p_{gas,0}, \cdots, p_{gas,n_v})
\ee
and is switched off otherwise. Here we use $\beta = 0.5$.
$|\triangle_x|(x_{\V}, y_{\V})$ and $|\triangle_y|(x_{\V}, y_{\V})$ are shorthand notations for absolute values of $\triangle_x$ and $\triangle_y$ at $(x_{\V}, y_{\V})$ respectively.

The second switch, \textbf{SW2},  which is used to pick out strong compressive motions
at the vicinity of the vertex $\V$, is accomplished by comparing the undivided divergence of the velocity to the smallest local signal speed. \textbf{SW2} is switched on if
\be
-\delta \min (C_0, \cdots, C_{n_v}) > (\triangle_x u_x + \triangle_y u_y)(x_{\V}, y_{\V})
\ee
and is switched off otherwise. Here we use $\delta = 0.1$.

On cell $\K_l$,
$$
C_l = \left( \frac{\gamma p_{gas,l}}{\rho_l} + \frac{\B_l\cdot \B_l}{\rho_l}\right)^{1/2}~, ~~~~l = 0, \cdots, n_v.
$$

When either \textbf{SW1} or \textbf{SW2} is switched on, it means that the region has a shock in it. In this case, we need to pick out the direction along which we want to upwind the evaluation of the electric field. We use a weighted combination to do that.

We estimate the direction $\mathbf{n}_{\rm S} = (n_{{\rm S},x}, n_{{\rm S},y})^T$ of the strong shock in the vicinity of the vertex $\V$ by
\begin{eqnarray}
n_{{\rm S},x} = \frac{\triangle_x p_{gas}}{\sqrt{ (\triangle_x p_{gas})^2 + (\triangle_x p_{gas})^2 } } \\
n_{{\rm S},y} = \frac{\triangle_y p_{gas}}{\sqrt{ (\triangle_x p_{gas})^2 + (\triangle_x p_{gas})^2 } }
\end{eqnarray}

Then for each $E_{z,l}$, we compute the associated weight $w_l$ by
\be
w_l = \frac{\alpha_l}{\sum^{n_v}_{s = 0} \alpha_s}
\label{eq:Ez_wei}
\ee
where $\alpha_l$ is defined by
\be
\alpha_l = (\mathbf{n}_{\rm S} \cdot \mathbf{n}_l)^4 + 10^{-6}~.
\ee
Here the small number $10^{-6}$ is to avoid division by zero in Eq. (\ref{eq:Ez_wei}).

We obtain
\be
E_z(x_{\V}, y_{\V}) = \sum^{n_v}_{l=0} w_l E_{z,l}(x_{\V}, y_{\V})~
\ee
at vertex $\V$ when discontinuity or strong compression is present.

\subsection{Time discretizations}

The method of lines approach is used to evolve the solution on the triangulated domain. Specifically, the third-order accurate TVD Runge-Kutta method \cite{ShuOsh88} is used to solve ordinary
differential equations (\ref{eq:hydro_scheme}) and (\ref{eq:semi_Bnor}).


\section{WENO-based  Reconstruction}
\label{sec:WENO}

The main ingredient of a high order accurate finite volume scheme is a reconstruction algorithm,
which reconstructs a smooth and high degree polynomial approximation of solutions from average values computed by the base finite volume scheme at the end of every Runge-Kutta stage. In return, the reconstructed polynomial is used for evaluating numerical fluxes in the subsequent calculation. In this section, we solve the following two sub-problems of reconstruction:

\smallskip

\noindent\emph{\textbf{Sub-problem 1}}. Given edge-length-averaged normal component $\overline{B_{n,j}}$ of $\Bxy$ define on cell edges $\mathcal{L}_j \in \mathfrak{E}_h $ and a positive integer $q$, for each cell $\K_i$, reconstruct an essentially non-oscillatory and divergence-free magnetic field $\widetilde{\B}^{xy}_{i}$
supported on $\K_i$.
Here $\widetilde{\B}^{xy}_{i} \in P_q(\K_i)^2$. $P_q(\K_i)$ is the space of polynomials of degree at most $q$ supported on $\K_i$. $\widetilde{\B}^{xy}_{i}$ is a $(q+1)^{th}$ order accurate approximation to exact $\Bxy$ (when it is smooth) on cell $\K_i$.
Moreover,
\be
\frac{1}{|\partial \K_{i,\ell}|} \int_{\partial \K_{i,\ell}}
\widetilde{\B}^{xy}_{i} \cdot \mathbf{n}_{i,\ell} dr =  \overline{B_{n,\ell_i^{-1}(\ell)}}~, ~~~~~\ell = 1, 2, 3.
\label{eq:B_consist}
\ee
Here $\partial \K_{i,\ell}$ is the $\ell^{th}$ edge of cell $\K_i$; $\mathbf{n}_{i,\ell}$ is the associated unit normal of this edge; $|\partial \K_{i,\ell}|$ is its length.
$\overline{B_{n,\ell_i^{-1}(\ell)}}$ is the edge-length-averaged normal component of the magnetic field on the cell edge $\partial \K_{i,\ell}$.
The local edge index $\ell$ and the global edge index $j$ is related by the mapping function (\ref{eq:edge_map}). We note that condition (\ref{eq:B_consist}) implies that
the piecewise smooth $\widetilde{\B}^{xy}_{i}$ agrees at the adjacent cell edges by
edge-length-averaged mean values. For our proposed scheme, we apply the reconstruction algorithm to solve this sub-problem at the end of every Runge-Kutta stage by using mean values $\overline{B_{h,n,j}}$, which is the numerical approximation of $\overline{B_{n,j}}$.

\bigskip

\noindent\emph{\textbf{Sub-problem 2}}.  Given cell average values $\overline{v}_i$ of a function $v(x,y)$  on each cell $\K_i$ and a positive integer $q$, for each cell $\K_i$, reconstruct an essentially non-oscillatory polynomial $\widetilde{P}_i(x,y)$ of degree
at most $q$ which has the mean value $\overline{v}_i$ and is a $(q+1)^{th}$ order accurate approximation to $v(x,y)$ on $\K_i$ (when $v(x,y)$ is smooth).
For our ideal MHD problem, at the end of every Runge-Kutta stage, this problem is solved for $\overline{v}_i$
replaced by the cell average values of the every component of $\U^H$ computed by the base finite volume scheme.

\smallskip

Before we describe the algorithm to solve these two reconstruction problems, we first recall several relevant concepts which will be used later in this section. See also \cite{KasIsk05} for details of related discussion. The level-0 von Neumann neighborhood of a triangle $\K \in \mathcal{T}_h$ contains the edge adjacent neighbors of $\K$ and  is defined to be the set
$$\mathfrak{N}^0(\K) = \left \{ \widetilde{\K} \in \mathcal{T}_h \setminus \{ \K \}: \widetilde{\K} \cap \K {\rm ~is ~an ~edge ~of}~ \K   \right \}~.$$
Here we neglect the subscript ``$i$'' of cells for convenience.
The level-$r$ von Neumann neighborhood is defined by the recursive definition
$$
\mathfrak{N}^r(\K) = \left (  \bigcup_{\widetilde{\K} \in \mathfrak{N}^{r-1}(\K) }
\mathfrak{N}^{r-1}(\widetilde{\K})   \right ) \setminus  \{ \K \}~,~~~~~ {\rm for}~~ r \geq 1~.
$$
For instance, the level-1 von Neumann neighborhood of $\K$ is by merging level-0 von Neumann neighborhoods of cells in $\mathfrak{N}^0(\K)$ which are edge adjacent neighbors of $\K$.

A critical component for the success of reconstruction on triangular meshes is the selection of a set of admissible stencils. Generally, this set should contain isotropic (or centered) stencil for achieving good approximation in smooth regions, and anisotropic (or one-sided and reverse-sided) stencils to avoid interpolation across
discontinuities \cite{KasIsk05,DumKas07}.

In order to construct these anisotropic stencils, the sector search algorithm \cite{KasIsk05,HarCha91} is utilized to construct forward sectors as well as backward sectors. Fig. \ref{fig:B_P1_1side} shows three forward sectors $FS_s, s = 1, 2, 3$ of cell $\K_0$. A forward sector is spanned by a pair of edges of $\K_0$.
Fig. \ref{fig:B_P1_rev} shows three backward sectors $BS_s, s = 1, 2, 3$ of $\K_0$. A
backward sector is defined by having its origin at the midpoint of a edge of $\K_0$ and its two boundary edges passing through the other two midpoints of remaining edges of $\K_0$.

When we perform a WENO reconstruction for a cell $\K \in \mathcal{T}_h$, in addition to construct a central stencil,  within each of the sectors of $\K$, we construct either an one-sided stencil (when it is a forward sector) or a reverse-sided stencil (when it is a backward sector). Additionally, when we construct an anisotropic stencil, we only include cells in von Neumann neighbors of $\K$, whose barycenters lie in the corresponding forward or backward sector.


\subsection{Divergence-free WENO-based reconstruction for $\Bxy$ on cell}
\label{sec:B_reconstruct}

Here we  describe a new divergence-free WENO-based reconstruction strategy for the magnetic field on triangular grids based on our recent work \cite{BalMey10} and \cite{Balsara04, BalRum09}. This solves the \emph{\textbf{Sub-problem 1}}.
We require that the reconstructed  $\widetilde{\mathbf{B}}_i^{xy}$
supported on cell $\K_i$ must satisfy the divergence-free condition on  $\K_i$ internally, is a
$(q+1)^{th}$ order accurate approximation to exact $\Bxy$ on $\K_i$ and also retains consistency at the cell boundaries in the  sense defined by Eq. (\ref{eq:B_consist}).
We summarize the reconstruction algorithm as follows:
\begin{itemize}
\item[Step 1.]  For every grid cell $\K_i$, we identify a
set of admissible reconstruction stencils
$\mathfrak{T}_{B} = \{ T_{B}^{(m)}: m = 1, \cdots,  7 \}$ using the method introduced in \cite{HarCha91,KasIsk05,DumKas07}. Here $T_{B}^{(1)}$ is the
central stencil;
$T_{B}^{(2)}$, $T_{B}^{(3)}$ and $T_{B}^{(4)}$ are the one-sided stencils  constructed in the forward sectors of $\K_i$;
and $T_{B}^{(5)}$, $T_{B}^{(6)}$ and $T_{B}^{(7)}$ are the reverse-sided stencils constructed in the backward sectors of $\K_i$ respectively.
 This choice of stencils allows to better limit oscillations of polynomial approximation of solutions supported on $\K_i$.

\item[Step 2.] We then use every stencil to reconstruct preliminarily a divergence-free magnetic
field $\B^{xy,(m)}_{i}, m = 1, \cdots, 7,$ with every component of $\B^{xy,(m)}_{i}$ represented by a polynomial function from cell edge-length-averaged values of normal component of the magnetic field $\mathbf{B}^{xy}$ defined on edges of cells contained in the stencil.

\item[Step 3.] For each preliminarily reconstructed $\B^{xy,(m)}_{i}$, a smoothness indicator $\omega_m$ is computed with $\sum^{7}_{m=1} \omega_m = 1$. The final nonlinearly stabilized WENO reconstruction $\widetilde\B^{xy}_{i}$ is defined by a weighted combination of $\sum^{7}_{m=1} \omega_m {\B}^{xy,(m)}_{i}$ and is exactly divergence-free.
\end{itemize}


\subsubsection{The second order accurate reconstruction for $\Bxy$}
\label{sec:B_P1}

\begin{figure}[!ht]
\subfigure[]{\label{fig:B_P1_cent}\epsfig{figure=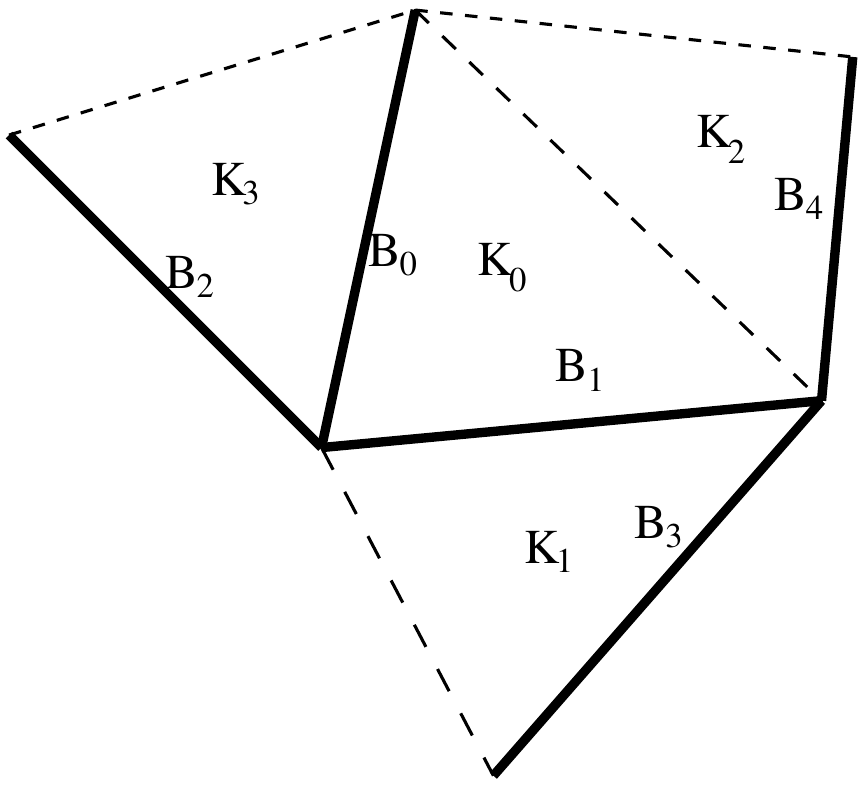,height=2.2in}}
\subfigure[]{\label{fig:B_P1_1side}\epsfig{figure=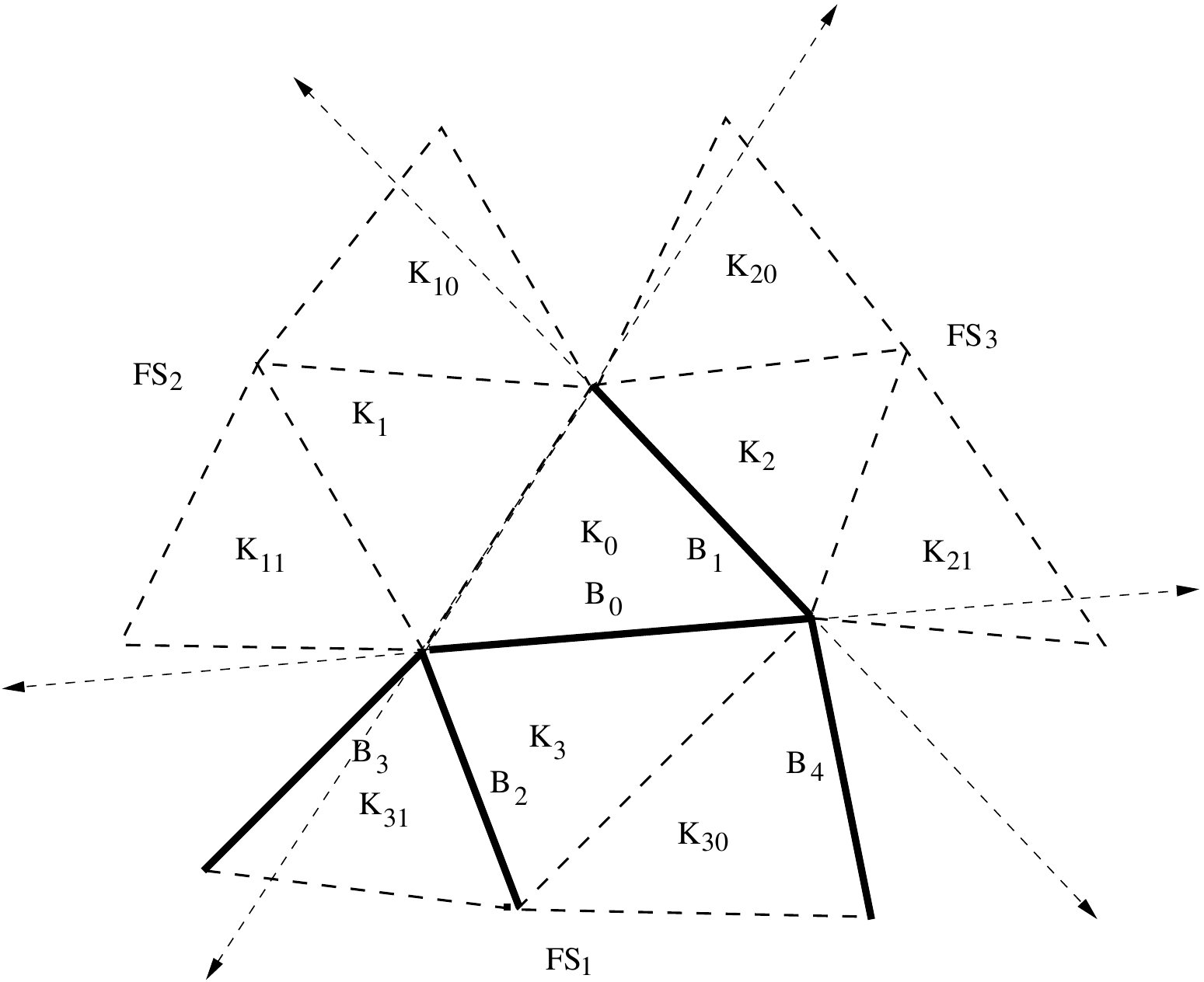,height=2.7in}}
\subfigure[]{\label{fig:B_P1_rev}\epsfig{figure=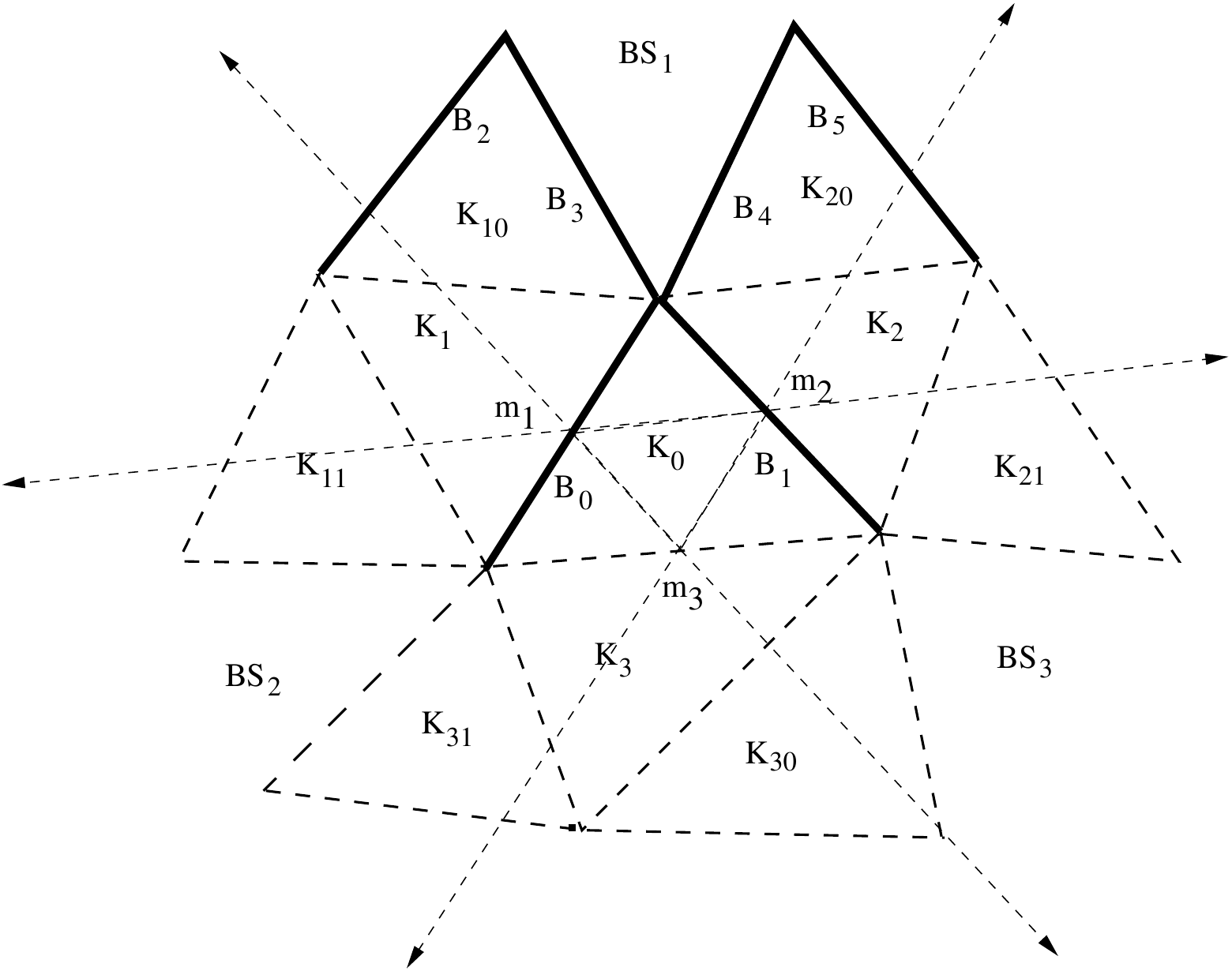,height=2.7in}}
\caption{Stencils for reconstructing second order accurate cell-centered divergence-free magnetic field on cell $\mathcal{K}_0$. Normal components of the magnetic field on solid line edges are utilized. (a) The central stencil.
(b) Three forward sectors $FS_1$, $FS_2$ an $FS_3$ of cell
$\K_0$ formed by spanning  a pair of edges of $\K_0$ respectively.
The cells of three one-sided stencils formed in each of the forward sector is shown here.
(c) Three backward sectors $BS_1$, $BS_2$ and $BS_3$ of $\K_0$.
A backward sector is defined by having its origin at the midpoint of an edge of $\K_0$ and its two boundary edges passing through the other two midpoints of remaining edges. The cells of three reverse-sided stencil formed  in each of the
backward sector is shown here.
\emph{Note that the same notations are used for different normal components of the magnetic field and cells in (a), (b)
 and (c) to avoid introducing too many notations.}
}
\label{fig:P1_B_field}
\end{figure}

To explain this reconstruction algorithm clearly, we first describe the second order accurate reconstruction algorithm. We note that our reconstructed $\widetilde{\mathbf{B}}_i^{xy}$ belongs to $ \{P_1(\mathcal{K}_i)^2,
\nabla \cdot \rBxy_i =0 \} $. The following linear polynomial expression is employed for representing
the preliminarily reconstructed $\B^{xy,(m)}=\left( B_x^{(m)}, B_y^{(m)} \right)^T$ as well as
$\widetilde{\mathbf{B}}_i^{xy}$
\be
\begin{array}{lll}
B_x^{(m)} (x, y) & =  & a_{0,m} + a_{1,m} x + a_{2,m} y ~, \\
B_y^{(m)}(x,y)  & = & b_{0,m} + b_{1,m} x + b_{2,m} y~.
\end{array}
\label{eq:P1_B_func}
\ee
Here we drop the subscript ``$i$'' of cells for convenience.
The
divergence-free condition $ \nabla \cdot \mathbf{B}^{xy,(m)} = 0$ gives the equation
\be
a_{1,m} + b_{2,m}  =  0 ~,
\label{eq:p1_div_cond}
\ee
for linear polynomial representation of $\mathbf{B}^{xy}$ by matching coefficients.
This reduces one of the degrees of freedom in preliminarily reconstructing  ${\mathbf{B}}^{xy,(m)}$.

Fig. \ref{fig:P1_B_field} shows  stencils  used to reconstruct $\widetilde{\B}^{xy}_0$ on cell $\K_0$.
The central stencil is shown in Fig. \ref{fig:B_P1_cent}.  Fig. \ref{fig:B_P1_1side} shows three forward sectors $FS_s, s = 1, 2, 3$ of $\K_0$ as well as cells used to construct corresponding one-sided stencils.
Fig. \ref{fig:B_P1_rev} shows three backward sectors $BS_s, s = 1, 2, 3$ of $\K_0$
and cells chosen to construct corresponding reverse-sided stencils.
The details of constructing one-sided and reverse-sided stencils in these sectors and
reconstructing linear polynomial $\widetilde{\B}^{xy}_0$ supported on $\K_0$ are given below.



\bigskip
\noindent{\bf Computation of the second order reconstruction polynomial for central stencil.}
\smallskip

See Fig. \ref{fig:B_P1_cent}. The central stencil consists of $\K_0$, and its three adjacent neighbors $\K_s, s = 1, 2, 3$, which are in the level-0 von Neumann neighbor of $\K_0$.

On $\K_0$, we arbitrarily choose two cell edge-length-averaged values of normal components of $\B^{xy}$, denoted by $B_0$ and $B_1$. On each of $\K_s$, $s = 1, 2, 3$,
we choose one cell edge-length-averaged value of normal components of $\B^{xy}$ defined on the cell edge which is not the common edge between $\K_0$ and $\K_s$, and denote these average values by $B_2$, $B_3$ and $B_4$ respectively.
The cell edges on which
these values are defined are relabeled by ${\mathcal{L}_s: s = 0, \cdots, 4}$ for convenience.
The solid lines in Fig. \ref{fig:B_P1_cent} indicate these edges.
We then solve
\be
\left\{
\begin{array}{lll}
a_{1,m} + b_{2,m} & = & 0 ~,\\
\frac{1}{|\mathcal{L}_s|}\int_{\mathcal{L}_s} \mathbf{B}^{xy,(m)}\cdot \mathbf{n}_s dr & = & {B}_{s}, ~~~s = 0, \cdots, 4~
\end{array}
\right .
\label{eq:P1_B_c}
\ee
to obtain a candidate ${\mathbf{B}}^{xy,(1)}$. Here $m = 0$; $|\mathcal{L}_s|$ is the length of
$\mathcal{L}_s$; and $\mathbf{n}_s$ is the unit normal of $\mathcal{L}_s$.

Notice that our choice of cell edges is such that we never form closed loops of edges. This is because that the divergence-free condition ensures that each closed loop has one redundant piece of information. Also notice that Eq. (\ref{eq:P1_B_func}) only has
five independent degrees of freedom and we have selected five  edges which do not close
from the central stencil. We remark that this general principle also applies to reconstructing $\Bxy$ on other stencils.

Finally we point out that for this central stencil, there are other possible ways to choose cell edges for reconstruction. See Fig. \ref{fig:B_P1_cent}. For instance, we could use values defined on the dashed line edges from every
neighboring cells of $\K_0$ together with ones defined on the solid line edges of $\K_0$ to reconstruct the polynomial. This is also the case for reconstruction by using other stencils. However, we notice that as long as we follow the principle of choosing edges close to $\K_0$, the results are not sensitive to choices of edges.


\bigskip
\noindent{\bf Computation of the second order reconstruction polynomials for one-sided stencils.}
\smallskip

Fig. \ref{fig:B_P1_1side} shows forward sectors and corresponding one-sided stencils utilized for reconstruction.
In Fig. \ref{fig:B_P1_1side}, cells $\mathcal{K}_1$, $\mathcal{K}_2$ and
$\mathcal{K}_3$  are  three neighbors of $\mathcal{K}_0$; and  $\mathcal{K}_{s0}$, $\mathcal{K}_{s1}$ are
two neighbors of  $\mathcal{K}_s$ (other than $\mathcal{K}_0$), $s=1, 2, 3.$
We form three one-sided stencils:
$T^{(2)}_B = \{\mathcal{K}_0,  \mathcal{K}_{3}, \mathcal{K}_{30}, \mathcal{K}_{31}\}$
in $FS_1$;
$T^{(3)}_B = \{\mathcal{K}_0,  \mathcal{K}_{1}, \mathcal{K}_{10}, \mathcal{K}_{11}\}$ in $FS_2$;
and
$T^{(4)}_B = \{\mathcal{K}_0,  \mathcal{K}_{2}, \mathcal{K}_{20}, \mathcal{K}_{21}\}$
in $FS_3$.

With every one-sided stencil $T^{(m)}_B, m = 2, 3, 4$, we will utilize cell edge-length-averaged values of normal component of the magnetic field $\mathbf{B}^{xy}$ on cell edges to  reconstruct preliminarily a divergence-free polynomial in the form of (\ref{eq:P1_B_func}).
Take stencil $T^{(2)}_B$ for example.
On $\mathcal{K}_0$, we utilize two average values of normal component of $\mathbf{B}^{xy}$ defined on its two edges respectively, and denote them by ${B}_{0}$ and ${B}_{1}$.
On $\mathcal{K}_3$, we use one average value of normal component of $\mathbf{B}^{xy}$ defined on the cell edge which is not the common edge between $\mathcal{K}_0$ and $\mathcal{K}_{3}$, and denote it by ${B}_{2}$. On each of $\mathcal{K}_{30}$ and $\mathcal{K}_{31}$, one average value defined on the cell edge which is not shared by $\mathcal{K}_1$ and $\mathcal{K}_{30}$ (or by $\mathcal{K}_3$ and $\mathcal{K}_{31}$) is employed respectively, and denote them by ${B}_{3}$ and
${B}_{4}$.


We now use $\{B_s : s = 0, \cdots, 4 \}$ to reconstruct preliminarily a piecewise linear $\widetilde{\mathbf{B}}^{xy,(2)}$
by solving  Eq. (\ref{eq:P1_B_c}). We similarly compute on the other two one-sided stencils to obtain two candidates respectively, denoted by ${\mathbf{B}}^{xy,(3)}$, and ${\mathbf{B}}^{xy,(4)}$.



\bigskip
\noindent{\bf Computation of the second order reconstruction polynomials for reverse-sided stencils.}
\smallskip

The reverse-sided stencils are constructed by using cells within the backward sectors.  Fig. \ref{fig:B_P1_rev} shows reverse-sided stencils which are constructed in the backward sector $BS_s, s = 1, 2, 3$ for reconstructing $\B^{xy}$ on $\mathcal{K}_0$.
Here cells $\mathcal{K}_1$, $\mathcal{K}_2$ and $\mathcal{K}_3$  are  three neighbors of $\mathcal{K}_0$; and  $\mathcal{K}_{s0}$, $\mathcal{K}_{s1}$ are
two neighbors of  $\mathcal{K}_s$ (other than $\mathcal{K}_0$), $s = 1, 2, 3.$
We construct three reverse-sided stencils:
$T^{(5)}_B = \{ \K_0, \K_{10}, \K_{20} \}$ in $BS_1$; $T^{(6)}_B = \{ \K_0, \K_{11}, \K_{31} \}$ in $BS_2$; and $T^{(7)}_B = \{ \K_0, \K_{21}, \K_{30} \}$ in $BS_3$.

Unlike the central or one-sided stencil case, here we employ a constrained least square method to solve the reconstruction problem by using each of the reverse-sided stencils.
Due to the divergence-free condition, in principle,
we only need 5 additional conditions to uniquely
determine functions (\ref{eq:P1_B_func}). However, a reverse-sided stencil can provide 6 admissible average values of normal component of $\B^{xy}$ (or 6 conditions to determine (\ref{eq:P1_B_func})). See Figure \ref{fig:B_P1_rev}. Take the stencil $T^{(5)}_B$ for example. The edges on which
the defined average values are employed for preliminary reconstruction are indicated by solid lines.  On $\K_0$, we can use two edge values $B_0$ and $B_1$. On $\K_{10}$, we can use $B_2$ and $B_3$; and on $\K_{20}$, we have
$B_4$ and $B_5$ to use. To avoid a bias in choosing values from $T^{(5)}_B$, we use all of these
6 average values and solve the following constrained least square problem for the preliminary reconstruction:
\be
\left \{
\begin{array}{lll}
\frac{1}{|\mathcal{L}_s|}\int_{\mathcal{L}_s} \mathbf{B}^{xy,(m)}\cdot \mathbf{n}_s dr & = & {B}_{s}, ~~~s = 2, ..., 5~, \\
\end{array}
\right .
\label{eq:B_P1_ls_ls}
\ee
~~~~~~~~~~~~~~~~~~~~~~~~~~~~~~~subject to:
\be
\left \{
\begin{array}{lll}
a_{1,m} + b_{2,m} & = & 0 ~,\\
\frac{1}{|\mathcal{L}_l|}\int_{\mathcal{L}_l} \mathbf{B}^{xy,(m)}\cdot \mathbf{n}_l dr & =&   {B}_{l}, ~~~l = 0,1~.
\end{array}
\right .
\label{eq:B_P1_ls}
\ee
When solving this constrained least square problem, Eq. (\ref{eq:B_P1_ls}) is satisfied exactly; while Eq. (\ref{eq:B_P1_ls_ls}) is satisfied in the least square sense. To improve the divergence-free aspect of the solution to this constrained least square problem on stencil cells other than $\K_0$, one can substitute Eq. (\ref{eq:p1_div_cond}) into Eq. (\ref{eq:B_P1_ls_ls}) before solving (\ref{eq:B_P1_ls_ls})-(\ref{eq:B_P1_ls}).
This ensures that the preliminarily reconstructed magnetic field on the cell of interest, $\K_0$,  is divergence-free, and matches the magnetic field defined on the cell edges bounding $\K_0$ by mean values exactly. Additionally, the preliminarily reconstructed magnetic field is the best approximation to the magnetic field on other cells in the reverse-sided stencil.

We denote candidates obtained by solving equations (\ref{eq:B_P1_ls})-(\ref{eq:B_P1_ls_ls})
in a constrained least square manner for every
reverse-sided stencils ${\mathbf{B}}^{xy,(5)}$, ${\mathbf{B}}^{xy,(6)}$ and
${\mathbf{B}}^{xy,(7)}$ respectively.

\bigskip
\noindent{\bf Computation of weights for second order WENO reconstruction.}
\smallskip

We now apply a weighted combination of  $\{ {\B}^{xy,(m)}: m = 1, \cdots,7\}$ to finalize $\widetilde{\B}_0^{xy}$ using the idea of WENO \cite{shu97a,Fri98}. Let
${\B}^{xy,(m)}$ be expressed by Eq. (\ref{eq:P1_B_func}).

We compute a quantity $\alpha_m$, which is the reciprocal of a smoothness measure by
$$\alpha_m = \frac{b_m}{\epsilon + ({a}_{1,m})^2 + ({a}_{2,m})^2 +
({b}_{1,m})^2 + ({b}_{2,m})^2}~, ~~~~~~~m = 1, \cdots, 7.
$$
Here we take $\epsilon = 10^{-6}$ to avoid division by zero. $b_m = 10$ when $m = 1$; and $b_m = 1$ otherwise. This follows the idea in \cite{LevPup99}.
The parameter $b_m$ allows to have more weight on the central stencil, which provides better accuracy when the solution is smooth.

The weight $\omega_m$ is computed by
$$\omega_m = \frac{\alpha_m}{\sum^7_{l=1} \alpha_l}~.
$$

Finally, we  reconstruct the piecewise linear polynomial approximation $\widetilde{\B}^{xy}_0$ on $\K_0$ by
$$
\widetilde{\mathbf{B}}^{xy}_0 = \sum^7_{m=1}\omega_m {\mathbf{B}}^{xy,(m)}.
$$
Notice that $\widetilde{\mathbf{B}}^{xy}_0$ satisfies the divergence-free condition: $\nabla \cdot \widetilde{\mathbf{B}}^{xy}_0 = 0$ exactly. This completes the second order accurate reconstruction for approximating $\Bxy$ on $\K_0$. In our proposed scheme, at the
end of every Runge-Kutta stage, we apply this reconstruction strategy by using
values $\overline{B_{h,n,j}}$ computed by the base finite volume scheme (\ref{eq:semi_Bnor}) to do the reconstruction.


\subsubsection{The third order accurate reconstruction for $\Bxy$}
\label{sec:B_P2}

\begin{figure}[!ht]
\subfigure[]{\label{fig:B_P2_cent}\epsfig{figure=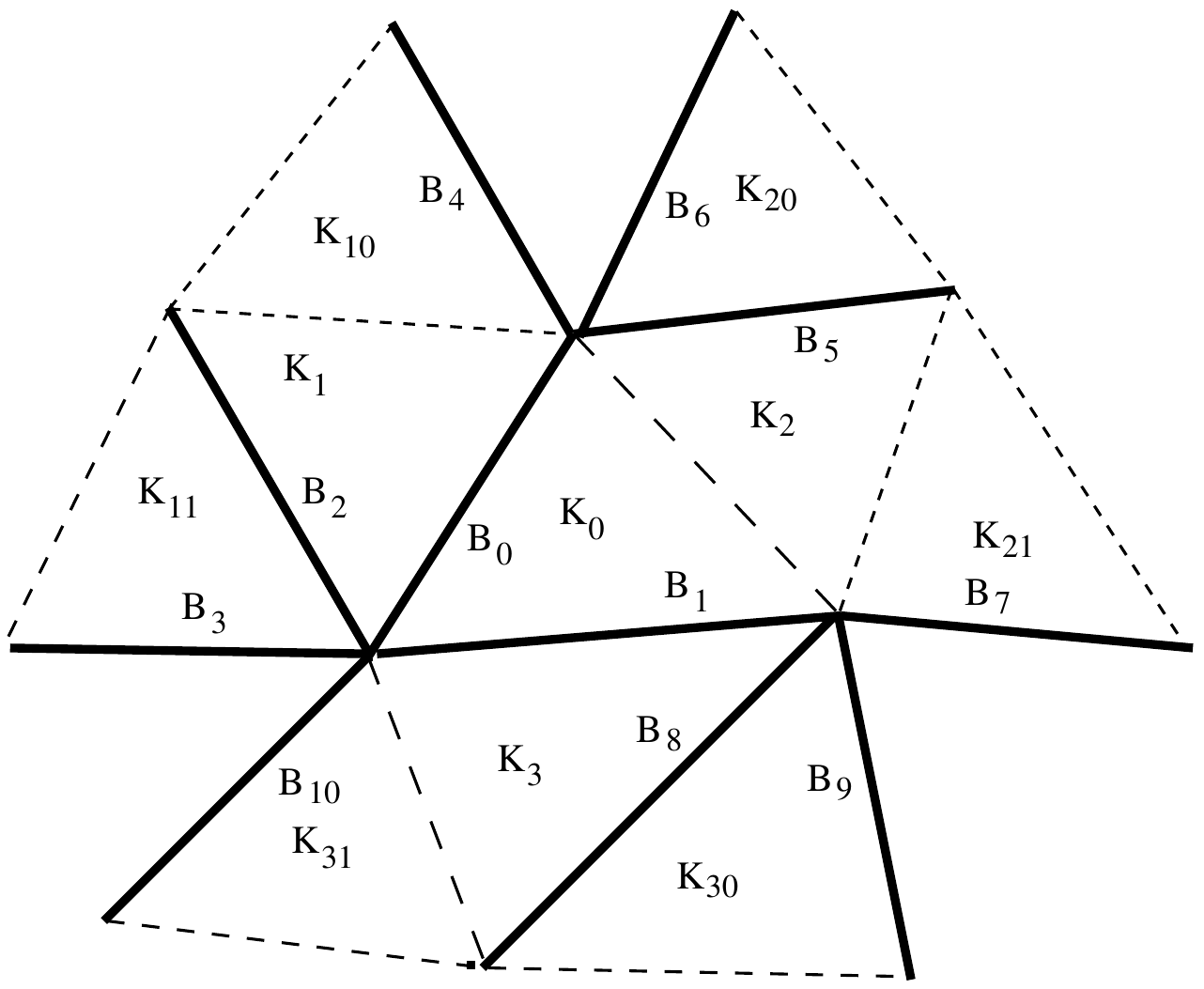,height=2.2in}}
\subfigure[]{\label{fig:B_P2_1side}\epsfig{figure=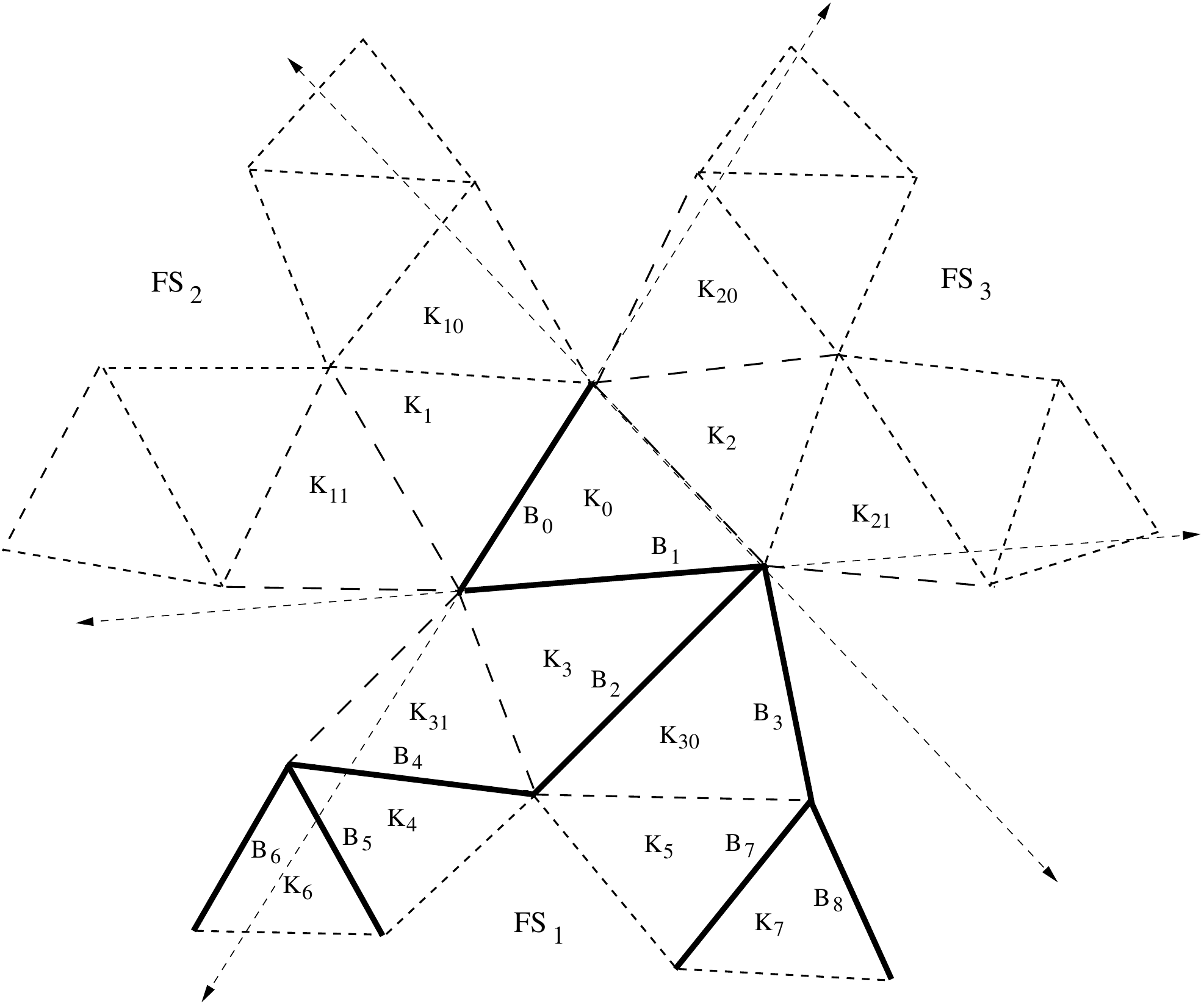,height=2.5in}}
\subfigure[]{\label{fig:B_P2_rev}\epsfig{figure=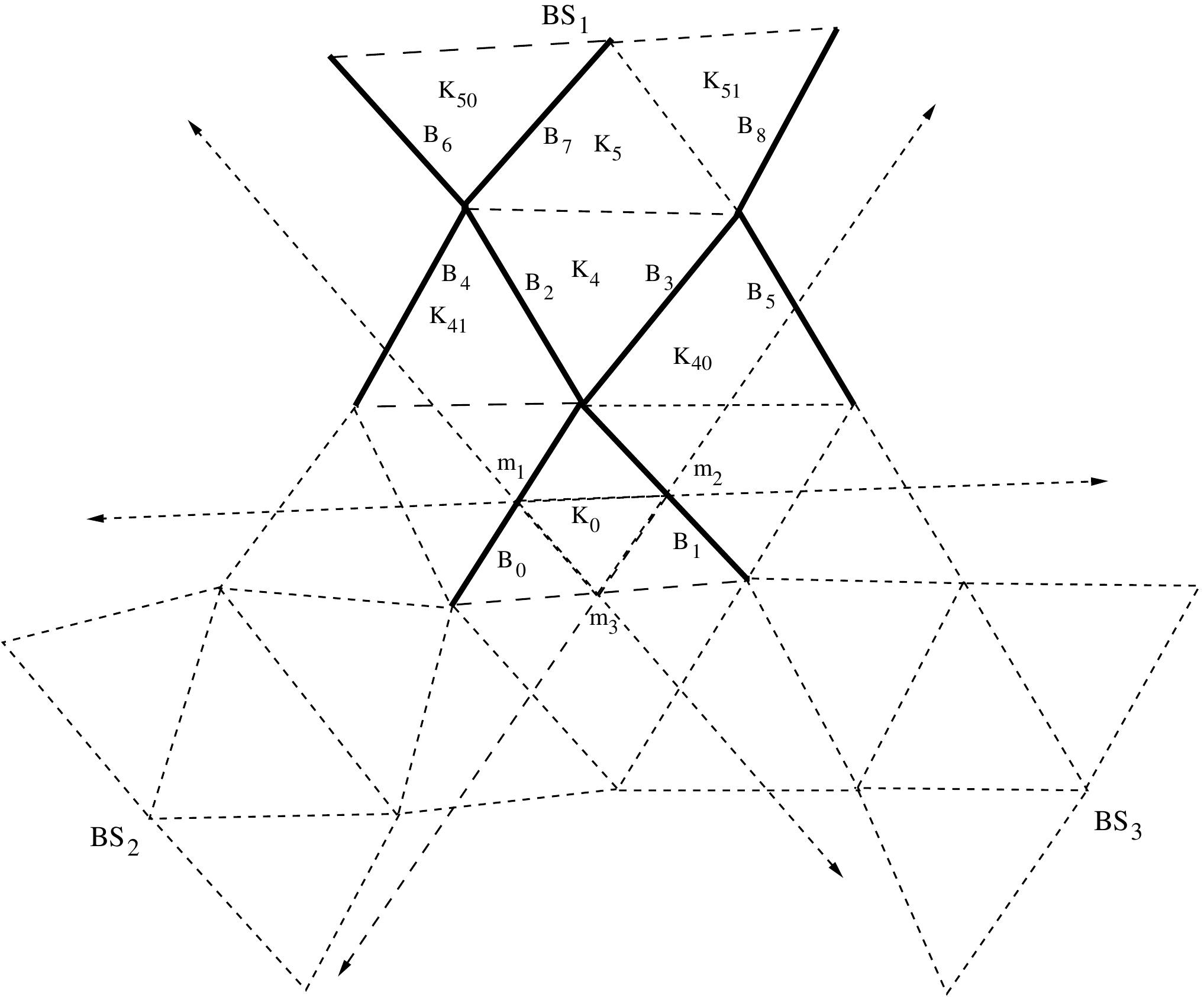,height=2.5in}}
\caption{Stencils for reconstructing the third order accurate cell-centered divergence-free magnetic field as well as for reconstructing the third
order accurate
polynomial approximations for cell centered variables on cell $\mathcal{K}_0$.
To reconstruct the divergence-free magnetic field, normal components of the magnetic field on solid line edges are utilized.
(a) The central stencil, (b) Three forward sectors $FS_1$, $FS_2$ an $FS_3$ of cell
$\K_0$ formed by spanning  a pair of edges of $\K_0$ respectively.
The cells of three one-sided stencils formed in each of the forward sectors  are shown here.
(c) Three backward sectors $BS_1$, $BS_2$ and $BS_3$ of $\K_0$. A backward sector is defined by having its origin at the midpoint of an edge of $\K_0$ and its two boundary edges passing through the other two midpoints of remaining edges of $\K_0$.
The cells of three reverse-sided stencils formed  in each of the
backward sectors are shown here.
\emph{Note that the same notations are used for different normal components of the magnetic field and cells in (a), (b)
 and (c) to avoid introducing too many notations.}
}
\label{fig:P2_B_field}
\end{figure}

The third order accurate divergence-free reconstruction of $\Bxy$ on cells
is a straight forward extension of the second order accurate reconstruction described in
Sec. \ref{sec:B_P1}. Thus we catalogue the key steps in implementing the third order accurate
case in this subsection.

The reconstructed $\widetilde{\mathbf{B}}^{xy}_i$ supported on cell $\K_i$, $\K_i \in \mathcal{T}_h$,
belongs to $ \{P_2(\mathcal{K}_i)^2,
\nabla \cdot \rBxy =0 \} $.
To avoid introducing too many notations, representation of the preliminarily reconstructed $\B^{xy,(m)} = \left(B_x^{(m)}, B_y^{(m)}\right)^T$ and  $\widetilde{\mathbf{B}}^{xy}_i$ is redefined by the following quadratic polynomial expression
\be
\begin{array}{lll}
B_x^{(m)} (x, y) & =  & a_{0,m} + a_{1,m} x + a_{2,m} y + a_{3,m} x^2 + a_{4,m} xy + a_{5,m} y^2~, \\
B_y^{(m)} (x,y)  & = & b_{0,m} + b_{1,m} x + b_{2,m} y + b_{3,m} x^2 + b_{4,m} xy + b_{5,m} y^2 ~.
\end{array}
\label{eq:P2_B_func}
\ee
The subscript ``$i$'' of cells is also dropped for convenience.
We note that the
divergence-free condition $ \nabla \cdot \mathbf{B}^{xy,(m)} = 0$ gives the following three equations
\be
\begin{array}{lll}
a_{1,m} + b_{2,m} & =  & 0 ~, \\
2a_{3,m} + b_{4,m}  & = & 0 ~, \\
a_{4,m} + 2b_{5,m} & = & 0
\end{array}
\label{eq:p2_div_cond}
\ee
for quadratic polynomial representation of $\mathbf{B}^{xy}$ by matching coefficients.
This reduces three of the degrees of freedom in preliminarily reconstructing
the third order accurate ${\B}^{xy,(m)}$.

Fig. \ref{fig:P2_B_field} shows stencils used for the third order accurate divergence-free reconstruction. The central stencil is shown in Fig. \ref{fig:B_P2_cent}. Fig. \ref{fig:B_P2_1side} shows
three forward sectors $FS_s, s = 1, 2, 3$ and cells in these forward sectors used to construct one-sided stencils. For the convenience, we still denote the one-sided stencil in $BS_1$ by $T^{(2)}_B$, the one in  $BS_2$ by $T^{(3)}_B$ and the one in  $BS_3$ by $T^{(4)}_B$. Fig. \ref{fig:B_P2_rev} shows three backward sectors $BS_s, s = 1, 2, 3$ and cells in these backward sectors used to construct corresponding reverse-sided stencils. Similarly, the reverse-sided stencil in $BS_1$ is denote by $T^{(5)}_B$,
the one in $BS_2$ by $T^{(6)}_B$ and the one in $BS_3$  by $T^{(7)}_B$.
 The details of constructing these stencils and
reconstructing polynomial $\widetilde{\B}^{xy}_i$ are given below.



\bigskip

\noindent{\bf Computation of the third order reconstruction polynomial for central stencil.}

\smallskip

See Fig. \ref{fig:B_P2_cent}. The central stencil $T_B^{(1)}$ of $\K_0$ contains cells in level-1 von Neumann neighbors of $\K_0$ and $\K_0$ itself.
On cell $\K_0$, we arbitrarily choose two cell edge-length-averaged values of normal components of $\Bxy$. On each of the remaining cells, we choose one edge-length-averaged values
on the edge which is connected to one of vertices of $\K_0$. We relabel these average values by $\{B_s: s = 0, \cdots, 10 \}$.
We then solve the following constrained least square problem to obtain $\B^{xy,(1)}$
\be
\left \{
\begin{array}{lll}
\frac{1}{|\mathcal{L}_s|}\int_{\mathcal{L}_s} \mathbf{B}^{xy,(m)}\cdot \mathbf{n}_s dr & = & {B}_{s}, ~~~s = 2,\cdots, 10~, \\
\end{array}
\right .
\label{eq:B_P2_c_ls}
\ee
~~~~~~~~~~~~~~~~~~~~~~~~~~~~~~subject to:
\be
\left\{
\begin{array}{lll}
a_{1,m} + b_{2,m} & =  & 0 ~, \\
2a_{3,m} + b_{4,m}  & = & 0 ~, \\
a_4 + 2b_5 & = & 0     \\
\frac{1}{|\mathcal{L}_s|}
\int_{\mathcal{L}_l} \mathbf{B}^{xy,(m)}\cdot \mathbf{n}_l dr & = & {B}_{l}, ~~~l = 0,1~.
\end{array}
\right .
\label{eq:B_P2_c}
\ee
Here $m=1$. We remark that when solving this constrained least square problem, Eq. (\ref{eq:B_P2_c}) is satisfied exactly; while Eq. (\ref{eq:B_P2_c_ls}) is satisfied in the least square sense. Similar to the second order accurate preliminary reconstruction on reverse-sided stencil, to improve the divergence-free aspect of the solution to this constrained least square problem on stencil cells other than $\K_0$,  Eq. (\ref{eq:p2_div_cond}) is substituted into Eq. (\ref{eq:B_P2_c_ls}) before solving (\ref{eq:B_P2_c_ls})-(\ref{eq:B_P2_c}).
This ensures that the preliminarily reconstructed magnetic field $\B^{xy,(1)}$ on $\K_0$ is
divergence-free, and matches the mean values of the magnetic field defined on the cell edges enclosing $\K_0$ exactly.


\bigskip

\noindent{\bf Computation of the third order reconstruction polynomials for one-sided stencils.}
\smallskip

Within each of the forward sectors, we construct an one-sided stencil.
See Fig. \ref{fig:B_P2_1side}.
Take the one-sided stencil $T_B^{(2)}$ constructed in the forward sector $FS_1$ for example. $T_B^{(2)}$ contains cells in level-3 von Neumann neighbors of $\K_0$,
which are close to the edges of $FS_1$,  and $\K_0$ itself. Thus
$T_B^{(2)} = \{ \K_0, \K_3, \K_{30}, \K_{31}, \K_4, \K_5, \K_6, \K_7 \}$.
One-sided stencils $T_B^{(3)}$ and $T_B^{(4)}$ are constructed similarly.

To reconstruct quadratic polynomial $\B^{xy,(2)}$ on $\K_0$ by using $T_B^{(2)}$, we choose two cell edge-length-averaged values of normal components of $\Bxy$ with one of them defined on the  edge shared by cells $\K_0$ and $\K_3$ (level-0 von Neumann neighbor). Then we choose $7$ edge values from the remaining cells.  The solid line edges shown in Fig. \ref{fig:B_P2_1side} give one admissible selection of these $7$ edge values.
We relabel these edge-length-averaged values by
$\{B_s : s = 0, \cdots, 8 \}$. Then  the following linear problem is solved to
reconstruct preliminarily a candidate $\B^{xy,(2)}$
\be
\left \{
\begin{array}{lll}
a_{1,m} + b_{2,m} & =  & 0 ~, \\
2a_{3,m} + b_{4,m}  & = & 0 ~, \\
a_{4,m} + 2b_{5,m} & = & 0     \\
\frac{1}{|\mathcal{L}_s|}\int_{\mathcal{L}_s} \mathbf{B}^{xy,(m)}\cdot \mathbf{n}_s dr & = & {B}_{s}, ~~~s = 0,\cdots, 8~. \\
\end{array}
\right .
\label{eq:B_P2_1s}
\ee
Here $m = 2$. The other two preliminarily  reconstructed polynomials $\mathbf{B}^{xy,(3)}$ by using $T_B^{(3)}$ and
 $\mathbf{B}^{xy,(4)}$ by using $T_B^{(4)}$ are computed in the same manner.


\bigskip

\noindent{\bf Computation of the third order reconstruction polynomials for reverse-sided stencils.}
\smallskip

See Fig. \ref{fig:B_P2_rev}. The reverse-sided stencils are constructed in the backward sectors respectively.  Take the reverse-sided stencil $T^{(5)}_B$ constructed in the backward sector $BS_1$ for example. $T^{(5)}_B$ contains two level-1 von Neumann neighbors $\K_{41}$ and $\K_{40}$, one level-2 von Neumann neighbor $\K_4$ which is adjacent to both $\K_{41}$ and $\K_{40}$, one level-3 von Neumann neighbor $\K_5$ which is adjacent to $\K_4$ and two level-4 von Neumann neighbors $\K_{50}$ and $\K_{51}$ which are adjacent to $\K_5$ (other than $\K_4$). The other two reverse-sided stencils $T^{(6)}_B$ and $T^{(7)}_B$ are constructed in the same manner.

To reconstruct the quadratic polynomial $\B^{xy,(5)}$ supported on $\K_0$ by using stencil $T^{(5)}_B$, we note that there are also multiple approaches to select cell edge-length-averaged values of normal component of $\Bxy$. Fig. \ref{fig:B_P2_rev} shows one choice.  Normal components of $\Bxy$ defined on the solid line edges in stencil $T^{(5)}_B$ are utilized. We choose two cell edge-length-averaged values of normal components of $\Bxy$ on two edges of $\K_0$ whose comment endpoint is in the
backward sector $BS_1$. We also choose four cell edge-length-averaged values of normal component of $\Bxy$ from edges of $\K_{40}$ and $\K_{41}$, which are level-1 von Neumann neighbors of $\K_0$. We then choose three  cell edge-length-averaged values of normal component of $\Bxy$ from edges of remaining cells in the stencil.

We relabel these edge-length-averaged values by
$\{B_s : s = 0, \cdots, 8 \}$. We then solve the linear problem (\ref{eq:B_P2_1s}) to
obtain $\B^{xy,(5)}$ The other two preliminarily reconstructed polynomials $\mathbf{B}^{xy,(6)}$ by using $T_B^{(6)}$ and $\mathbf{B}^{xy,(7)}$ by using $T_B^{(7)}$ are computed in the same manner.

\bigskip
\noindent{\bf Computation of weights for third order WENO reconstruction.}
\smallskip

We now apply a weighted combination of  $\{{\B}^{xy,(m)}: m = 1, \cdots,7\}$ to finalize $\widetilde{\B}^{xy}_0$ using the idea of WENO \cite{shu97a,Fri98}. Let $\mathbf{B}^{xy,(m)} = \left( B_x^{(m)}(x,y), B_y^{(m)}(x,y) \right)^T$ be expressed by Eq. (\ref{eq:P2_B_func}).

We first compute smoothness measures of $x$-component and $y$-component of $\mathbf{B}^{xy,(m)}$ respectively by
\be
SI({B}^{(m)}) = \left( \sum_{|\beta| = 1} \int_{\K_0} h^{-2} (D^{\beta} B^{(m)}(x,y) )^2dxdy   \right)^{1/2}~.
\ee
Here $B$ stands for either $B_x^{(m)}(x,y)$ or $B_y^{(m)}(x,y)$.

The $\alpha_m$ is redefined by
$$\alpha_m = \frac{b_m}
{\left( \epsilon + SI({B_x}^{(m)}) + SI({B_y}^{(m)}) \right)^4}~,
~~~~~~~m = 1, \cdots, 7.
$$
$\epsilon = 10^{-6}$ is used to avoid division by zero. $b_m = 10$ when $m = 1$; and $b_m = 1$ otherwise \cite{LevPup99}.

The weight $\omega_m$ for the third order accurate divergence-free reconstruction is computed by
$$\omega_m = \frac{\alpha_m}{\sum^7_{l=1} \alpha_l}~.
$$

Finally, we  reconstruct the piecewise quadratic polynomial approximation $\widetilde{\B}^{xy}_0$ on $\K_0$ by
$$
\widetilde{\mathbf{B}}^{xy}_0 = \sum^7_{m=1}\omega_m {\mathbf{B}}^{xy,(m)}.
$$
Notice that $\widetilde{\B}^{xy}_0$ satisfies the divergence-free condition $\nabla \cdot \widetilde{\B}^{xy}_0 = 0$ exactly. This completes the third order accurate divergence-free reconstruction for the magnetic field on $\K_0$. Again, at the
end of every Runge-Kutta stage, we apply this reconstruction method by using
values $\overline{B_{h,n,j}}$ computed by the base finite volume scheme (\ref{eq:semi_Bnor}) to do the reconstruction.

\subsection{WENO finite volume reconstruction for $\U^H$}
\label{sec:WENO_cell}

Here we describe an algorithm to reconstruct polynomials of degree $q = 3$ from
given cell averages to solve the \emph{\textbf{Sub-problem 2}}.

Let $(x_i, y_i)$ be the coordinates of the barycenter of cell  $\mathcal{K}_i$. We use the following monomial expression of a second degree polynomial $P_i(x,y)$ supported on $\K_i$:
\be
{P}_i(x,y) = a_{0,i} + a_{1,i}(x-x_i) + a_{2,i}(y-y_i) + a_{3,i}(x-x_i)^2 +
 a_{4,i}(x-x_i)(y-y_i) + a_{5,i} (y-y_i)^2~.
\ee
To reconstruct a polynomial function approximation to a function $v(x,y)$ on cell $\mathcal{K}_i$ from cell average values $\overline{v}_i$ of $v(x,y)$, we also follow the
reconstruction algorithm  described in Sec. \ref{sec:B_reconstruct} except that in Step 2 of the algorithm,
we use cell average values here for solving the  \emph{\textbf{Sub-problem 2}}; and we do not require the reconstructed  $\widetilde{P}_i(x,y)$ to be divergence-free.

For the self-completeness of the paper,  we briefly describe the WENO reconstruction of the second degree polynomial $\widetilde{P}_0(x,y)$ on cell $\K_0$. We refer to \cite{DumKas07,Fri98,KasIsk05} for description of the first degree polynomial reconstruction.
See Fig. \ref{fig:P2_B_field} for all stencils used in the second degree polynomial reconstruction.
On each of the stencil $T^{(m)}, m = 1, \cdots, 7$,  we first  reconstruct preliminarily a polynomial $P_{0}^{(m)}(x,y)$  supported on $\K_0$ by solving a system of linear equations (or a constrained least square problem) respectively.

The central stencil $T^{(1)}$ to reconstruct preliminarily a second degree polynomial
${P}_{0}^{(1)}(x,y)$ on cell $\K_0$ is shown in Fig. \ref{fig:B_P2_cent}.
$T^{(1)}$ consists of cell $\K_0$, its three neighbors $\K_1$, $\K_2$ and $\K_3$, and
$\K_{s0}$ and $\K_{s1}$ which are two neighbors of $\K_s$ (other than $\K_0$), $s =1, 2, 3$. Thus $T^{(1)} = \{\K_0, \K_s, \K_{s0}, \K_{s1}, s = 1, 2, 3\}$, which consists of  the level-1 von Neumann neighbors of $\K_0$ and $\K_0$ itself.
The coefficients of ${P}_{0}^{(1)}(x,y)$ are determined by solving the following
constrained linear problem
\be
\begin{array}{l}
\left \{
\begin{array}{lll}
\int_{\K_{s l}} P_0^{(1)}(x,y) dxdy & = & |\K_{s l}| \overline{v}_{s l}~,
~~~~s = 1, 2, 3; ~~~l = 0, 1; \\
\end{array}
\right .
\\
{ \textrm{subject ~to:} }
 \\
\left\{
\begin{array}{lll}
\int_{\K_r} P_0^{(1)}(x,y) dxdy & = & |\K_r| \overline{v}_r~,
~~~~r = 0, 1, 2, 3;
\end{array}
\right .
\end{array}
\ee
where  $|\K|$ is the area of cell $\K$; $\overline{v}_{s l}$ is the cell average value defined on cell
$\K_{s l}$; and $\overline{v}_r$ is the cell average value defined on $\K_r$.

Within in three forward sectors $FS_s, s = 1, 2, 3$, we construct three
one-sided stencils $T^{(2)}$, $T^{(3)}$ and $T^{(4)}$ respectively.  See Fig. \ref{fig:B_P2_1side}. In $FS_1$,
$T^{(2)} = \{ \K_0, K_3, K_{30}, \K_{31}, \K_4, \K_5 \}$.
Thus $T^{(2)}$  consists of $\K_0$, level-1 von Neumann neighbors of $\K_0$ in this sector, and two additional cells $\K_4$ and $\K_5$  which are neighbors of level-1 von Neumann neighbors in this sector (other than $\K_3$). The other two one-sided stencils $T^{(3)}$  and $T^{(4)}$  are constructed similarly.

Then by using every stencil $T^{(m)}, m = 2, 3, 4$, we reconstruct preliminarily polynomials
${P}_{0}^{(m)}(x,y)$ respectively  by solving the following linear system
\be
\int_{\triangle_s \left(T^{(m)} \right)} P_0^{(m)}(x,y) dxdy = |\triangle_s(T^{(m)})| \overline{v}_s~,
~~~~s = 1, \cdots, 6,
\ee
where $\triangle_s(T^{(m)}) \in T^{(m)}$ is a cell in $T^{(m)}$, $m = 2, 3, 4$; $\overline{v}_s$ is
the cell average defined on $\triangle_s(T^{(m)})$, and $|\triangle_s(T^{(m)})|$ is the cell area
of $\triangle_s(T^{(m)})$.


To further improve the robustness of the scheme, the reverse-sided stencils are also included. See Fig. \ref{fig:B_P2_rev}. Within three backward sectors $BS_s, s = 1, 2, 3$,
we construct three reverse-sided stencils $T^{(5)}$, $T^{(6)}$ and $T^{(7)}$ respectively.

In $BS_1$, $T^{(5)} = \{ \K_0, \K_4, \K_{40}, \K_{41}, \K_5, \K_{50}, \K_{51} \}$.  The method to construct this stencil is the same as the one used to
construct the stencil $T_B^{(5)}$ used to  reconstruct a third order accurate divergence-free magnetic field $\B^{xy,(5)}$ described in Sec. \ref{sec:B_P2}.
Stencils $T^{(6)}$ and $T^{(7)}$ are constructed similarly.

We next reconstruct preliminarily polynomial $P_0^{(5)}$ by using
stencil $T^{(5)}$  by solving the following constrained least square problem
\be
\begin{array}{l}
\left \{
\begin{array}{lll}
\int_{\K_{l}} P_0^{(5)}(x,y) dxdy & = & |\K_{l}| \overline{v}_l~,
 ~~~l = 4, 5, 40, 41, 50, 51; \\
\end{array}
\right .
\\
{ \textrm{subject ~to:} }
 \\
\left\{
\begin{array}{lll}
\int_{\K_0} P_0^{(5)}(x,y) dxdy & = & |\K_0| \overline{v}_0~;
\end{array}
\right .
\end{array}
\ee

Polynomials ${P}_{0}^{(6)}(x,y)$ reconstructed by using $T^{(6)}$  and
${P}_{0}^{(7)}(x,y)$ reconstructed by using  $T^{(7)}$ are computed similarly.


For each $P_0^{(m)}(x,y), m = 1, \cdots, 7$, we compute a smoothness indicator \cite{Fri98} by
\be
SI(P_0^{(m)}) = \left( \sum_{|\alpha=1|} \int_{\K_0}  h^{-2} (D^\alpha P_0^{(m)}(x,y))^2 dxdy  \right)^{1/2}~.
\ee
This smoothness indicator is suitable for stringent shock wave interaction problems. See \cite{Fri98} for discussion of other oscillation indicators.

Weights $\omega_m$ from these smoothness indicators are redefined by
\be
\omega_m = \frac{b_m \left(\epsilon +  SI(P_0^{(m)}) \right)^{-4} }
{ \sum^7_{\ell=1} b_m \left(\epsilon +  SI(P_0^{(\ell)}) \right)^{-4}    }~,
\ee
where $b_m = 10$ when $m = 1$ and $b_m = 1$ otherwise.
$\epsilon = 10^{-6}$ is used to avoid division by zero.

The final nonlinear WENO reconstruction polynomial $\widetilde{P}_0(x,y)$ is defined by
\be
\widetilde{P}_0(x,y) = \sum^7_{m=1} \omega_m P_0^{(m)}(x,y) ~.
\ee
This completes the reconstruction for approximating $v(x,y)$ on $\K_0$.
In the present paper, this reconstruction algorithm is applied at the end of every Runge-Kutta stage, and used to reconstruct every component of
$\U^H$ with $\overline{v}_i$ replaced by the cell average values of corresponding component of
$\U^H$ computed by the base finite volume scheme.

\section{Numerical Test Problems}
\label{sec:test}

\subsection{Vortex evolution problem}
\label{sec:vort}

We consider a vortex evolution problem, which was initially
suggest in \cite{shu97a} and
was adapted to the MHD equations in \cite{Balsara04}, to assess the convergence order of the scheme.

The problem is defined on a $[-5, 5] \times [-5, 5]$ domain with periodic boundary conditions on both sides.  The unperturbed MHD flow is given by
$(\rho, p_{gas}, u_x, u_y, B_x, B_y) = (1, 1, 1, 1, 0, 0)$. The ratio of specific heats
is $\gamma = 5/3.$
The vortex is introduced through perturbed velocity and magnetic fields given
by
$$
(\delta u_x, \delta u_y) = \frac{\kappa}{2\pi} e^{0.5(1-r^2)}(-y, x)~,~~~~~~~~~
(\delta B_x, \delta B_y) = \frac{\mu}{2\pi} e^{0.5(1-r^2)}(-y, x)~,
$$
where $r^2 = x^2 + y^2.$
The pressure determined by the dynamical balance is
given by
$$
\delta p_{gas} = \frac{\kappa^2(1-r^2) - \mu^2}{8 \pi^2} e^{1-r^2}.
$$
We use $\kappa = 1$, $\mu = 1$ in our computation. The exact solution is the initial configuration propagating with speed $(1,1)$,
and is given by
$$
U(x,y,t) = U_0(x-t,y-t).
$$
The computation domain is $[-5, 5] \times [-5, 5]$. Periodic boundary condition is used at both side of the domain. The periodic boundary condition introduces an error of
magnitude $\mathcal{O}(10^{-6})$, which does not affect the reported results. The typical triangle edge length, denoted by $h$, is listed in the first column of all the tables shown in this section. We show the $L_1$ and $L_\infty$ errors and orders of accuracy of variables $\rho$ and $\varepsilon$ at time $T = 1.0$.
Table \ref{tab:P1FV_Vort} shows  the result for the second order accurate
divergence-free WENO reconstruction-based scheme. The result for the third
order accurate scheme is shown in Table \ref{tab:P2FV_Vort}.
As can be seen, results in these two tables show clearly that we have achieved the expected accuracy property of the scheme. The absolute value of the undivided divergence of the magnetic field is about $\mathcal{O}(10^{-13})$ in
these simulations.

\begin{table}[h!b!p!]
 \caption{Numerical errors and convergence order for the second order accurate divergence-free WENO reconstruction-based method for solving the 2D vortex evolution problem.}
\begin{center}
\begin{tabular}{|l|l|l|l|l|l|l|l|l|}
\hline
h & {$L_1$ }& order & {$L_\infty$} & order  &
{$L_1$}& order & {$L_\infty$} & order \\
& $\rho$ error & &  $\rho$ error & &  $\varepsilon$ error & &   $\varepsilon$ error & \\
\hline
1/40 &  2.93E-3 & -  & 7.65E-3  & - & 1.07E-1 & - & 3.69E-2 & -  \\
\hline
1/80 &  8.22E-3 & 1.83 & 4.61E-3 & 0.73 & 2.71E-2 & 1.98 & 1.42E-2 &  1.38 \\
\hline
1/160 & 1.38E-3 & 2.57 & 1.31E-3 & 1.81 & 4.95E-3 & 2.46   &  4.84E-3 &  1.55 \\
\hline
1/320 & 2.40E-4 & 2.53 & 3.92E-4 &  1.74 & 9.23E-4 & 2.42  &  1.71E-3 &  1.49 \\
\hline
\end{tabular}
\end{center}
\label{tab:P1FV_Vort}
\end{table}

\begin{table}[h!b!p!]
 \caption{Numerical errors and convergence order for the third order accurate divergence-free WENO reconstruction-based method for solving the 2D vortex evolution problem.}
\begin{center}
\begin{tabular}{|l|l|l|l|l|l|l|l|l|}
\hline
h & {$L_1$ }& order & {$L_\infty$} & order  &
{$L_1$}& order & {$L_\infty$} & order \\
& $\rho$ error & &  $\rho$ error & &  $\varepsilon$ error & &   $\varepsilon$ error & \\
\hline
1/20 &  2.01E-2 & -  & 2.80E-3  &  - & 1.31E-1 & - & 2.49E-2 & - \\
\hline
1/40 &  2.31E-3 & 3.12  & 4.96E-4  & 2.50 & 1.03E-2 & 3.67 & 2.44E-3 & 3.36  \\
\hline
1/80 &  1.85E-4 & 3.64 & 4.44E-5 & 3.48 & 6.64E-4 & 3.96 & 2.03E-4 &  3.59 \\
\hline
1/160 & 2.58E-5 & 2.84 & 8.75E-6 & 2.34 & 9.26E-5 & 2.84   &  1.90E-5 &  3.42 \\
\hline
\end{tabular}
\end{center}
\label{tab:P2FV_Vort}
\end{table}

\old{
\subsection{Smooth ${\rm Alfv\acute{e}n}$ wave problem}
\label{sec:smooth}

We consider a smooth ${\rm Alfv\acute{e}n}$ wave problem from \cite{Tot00}
to check the accuracy of the proposed scheme. This problem is the
propagation of a circularly polarized ${\rm Alfv\acute{e}n}$ wave in the
domain $[0, 1/\cos\alpha] \times [0, 1/\sin\alpha]$, where $\alpha$ is the wave
propagation angle relative to the x-axis. We use $\alpha = \pi/4$ in our test.

The initial conditions are taken as
$$
\rho = 1 ~; ~~~ u_{\|} = 0 ~; ~~~ u_{\bot} = 0.1\sin(2\pi\xi) ~;~~~ u_z = 0.1\cos(2\pi\xi) ~;
$$
$$
p=0.1 ~;~~~ B_{\|} = 1 ~; ~~~ B_{\bot} = u_{\bot} ~; ~~~ B_z = u_z~;
$$
where $\xi = x\cos(\alpha) + y\sin(\alpha).$

For this test problem, the ${\rm Alfv\acute{e}n}$ wave propagates
periodically towards the origin with a constant ${\rm Alfv\acute{e}n}$ wave
speed $B_{\|}/\sqrt \rho = 1$ and returns to its initial state when the time $t$ is an integer.

Following \cite{Tot00}, we estimate the relative numerical error of variable $u$ by
$$
\delta(u) = \frac{\sum_{i=1}^{\mathcal{N}} |u_i - u_i^{\rm exact}|}
{\sum^{\mathcal{N}}_{i=1}|u_i^{\rm exact}| }~,
$$
where $u_i$ is the numerical solution at the centroid of the $i^{th}$ cell.

The average numerical errors are defined by
$$
\delta = \frac{1}{4}(\delta(u_{\bot}) + \delta(u_z) + \delta(B_{\bot}) + \delta(B_z) )~,
$$
}

\subsection{Numerical dissipation and long-term decay of ${\rm Alfv\acute{e}n}$ waves}
\label{sec:decay}
We consider a smooth solution problem proposed in \cite{Balsara04}, which examines the numerical dissipation of torsional ${\rm Alfv\acute{e}n}$ waves that are made to propagate at a small angle to the y-axis.
We use the same angle $\alpha = \tan(1/6) = 9.462^0$; and the magnetic field is normalized by a $1/\sqrt{4\pi}$ factor. The density $\rho_0 = 1$, and
pressure $p_0 = 1$ are initial values of density and pressure respectively.
The unperturbed velocity is $u_0 = 0$, and the unperturbed magnetic field is $B_0 = 1$.

The computational domain is $[-r/2,r/2]\times [-r/2,r/2]$ with $r=6$. The direction of wave propagation is along the unit vector
$
\mathbf{n} \equiv (n_x, n_y) = (\frac{1}{\sqrt{r^2 + 1}}, \frac{r}{\sqrt{r^2 + 1}} )~.
$
The phase of the wave is taken to be
$
\phi = \frac{2\pi}{n_y}(n_x x + n_y y - V_A t)~,
$
where $V_A = B_0\sqrt{\rho_0}$.
The velocity is given by
$
\vel = (u_0 n_x-\epsilon n_y\cos \phi, ~ u_0 n_y + \epsilon n_x \cos \phi, ~\epsilon \sin \phi)~,
$
where $\epsilon = 0.2$.
The magnetic field is given by
$
\B = (B_0 n_x + \epsilon n_y \sqrt{\rho_0} \cos \phi,~  B_0 n_y - \epsilon n_x \sqrt{\rho_0} \cos \phi, ~ -\epsilon \sqrt{\rho_0} \sin \phi )~.
$

\begin{figure}[ht]
\begin{center}
\subfigure[]{\label{fig:decay_Vz}\epsfig{figure=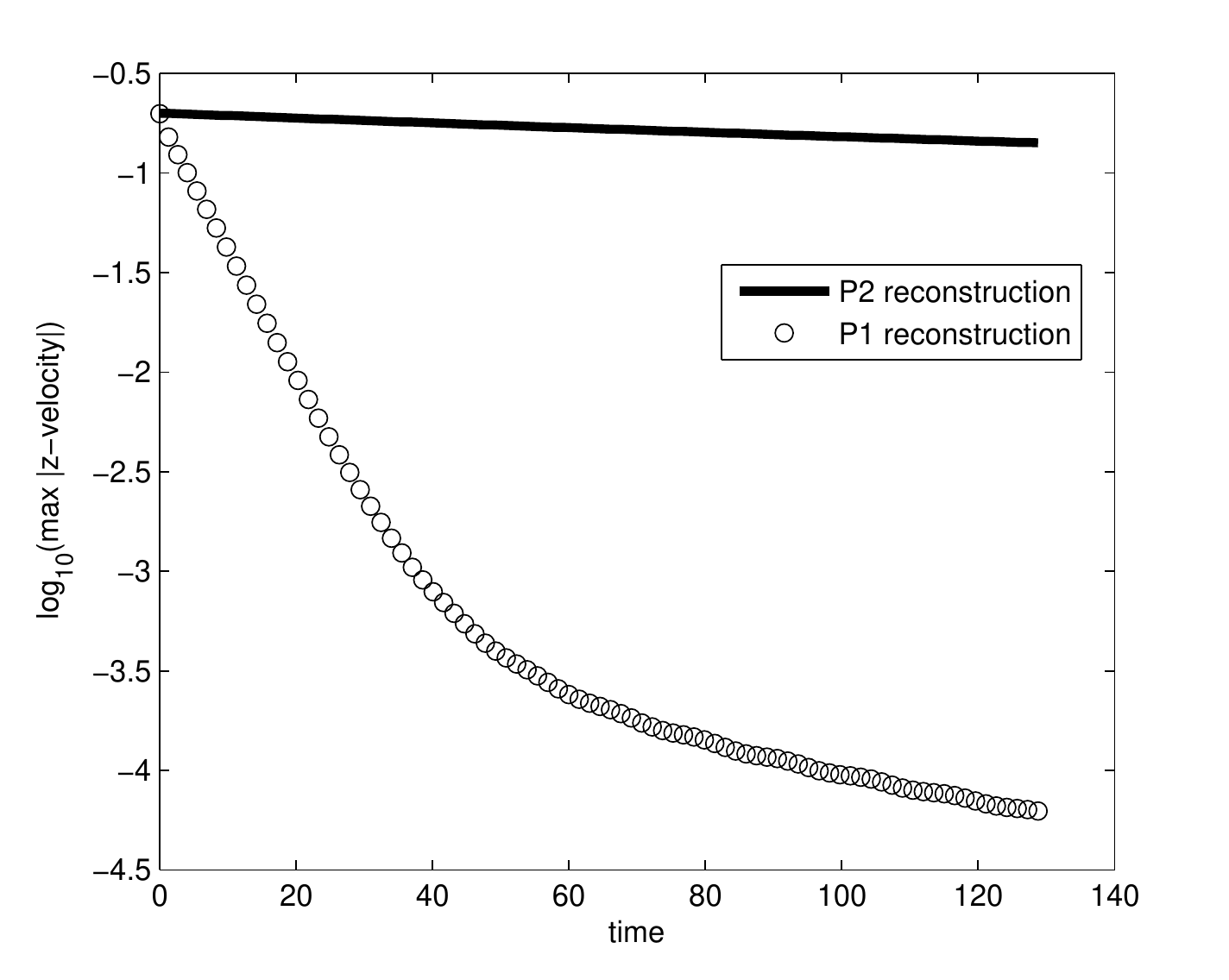,height=2.4in}}
\subfigure[]{\label{fig:decay_Bz}\epsfig{figure=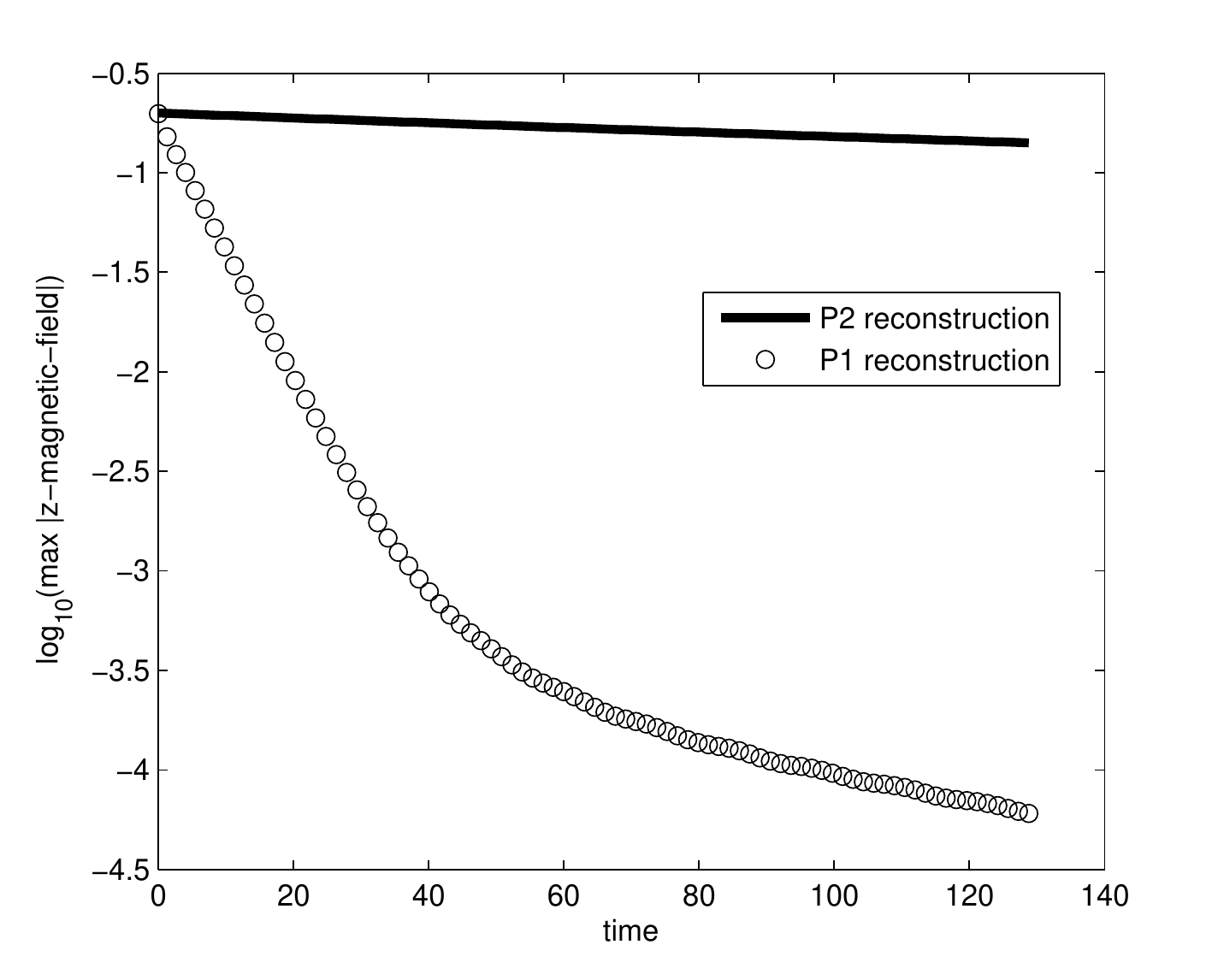,height=2.4in}}
\caption{(a) Logarithm plot of the maximum of absolute value of the $z$-component of the velocity
from the torsional  ${\rm Alfv\acute{e}n}$ waves propagation problem.
(b) Logarithm plot of the maximum of the absolute value of the $z$-component of the magnetic field
from the torsional  ${\rm Alfv\acute{e}n}$ waves propagation problem.
The maximum value should remain constant for the exact
solution but decay due to the numerical dissipation. The solid line
is for the third order accurate scheme; while the circled line is
for the second order accurate scheme.
\label{fig:decay_alfven}
}
\end{center}
\end{figure}

The computational domain is $[-3, 3] \times [-3, 3]$. The typical edge length of triangles is roughly equal to $\frac{1}{20}$. Solution of the problem is computed to a time $T = 129$. The maximum values of $u_z$ and $B_z$ should remain constant over time for the exact solution, but decay due to the numerical dissipation. Therefore this problem provides a good assessment of dissipation of the numerical scheme.
Figure \ref{fig:decay_alfven} shows the logarithm of the maximum of absolute values of $u_z$ and $B_z$ over time. We see clearly  that the third order accurate scheme is substantially less dissipative than the second order accurate scheme.

In what follows, we test the problems with discontinuities to assess
the non-oscillatory property of the proposed third order accurate scheme.

\subsection{Rotor problem}
\label{sec:2nd_rotor}
This test problem is first proposed in \cite{BalSpi99b} and is considered as the second rotor problem in \cite{Tot00}. The computational domain is $[0, 1] \times [0, 1]$.
$\gamma = 5/3.$ A dense rotating disk of fluid is initially placed at the central area of the computational domain, while the ambient fluid is at rest.
The initial condition is given by
$$
(\rho, p_{gas}, u_x, u_y, u_z, B_x, B_y, B_z) =
(\rho(\bx,0), 0.5, u_x(\bx,0), u_y(\bx,0), 0, \frac{2.5}{\sqrt{4\pi}}, 0, 0, 0).
$$
Here
$$
(\rho(\bx,0), u_x(\bx,0), u_y(\bx,0) )= \left\{
\begin{array}{llll}
10, & -(y-0.5)/r_0, & (x-0.5)/r_0 & {\rm if} ~r < r_0 \\
1+ 9f, & -(y-0.5)f/r, &  (x-0.5)f/r & {\rm if} ~r_0 < r < r_1 \\
1, & 0, &0, & {\rm if} ~r > r_1
\end{array}
\right .
$$
where $r_0 = 0.1$, $r_1 = 0.115$, $f = (r_1-r)/(r_1-r_0)$,
and $r = \sqrt{(x-0.5)^2 + (y-0.5)^2}$.

\begin{figure}[!ht]
\centering
\subfigure[]{\label{fig:den_rotor}\epsfig{figure=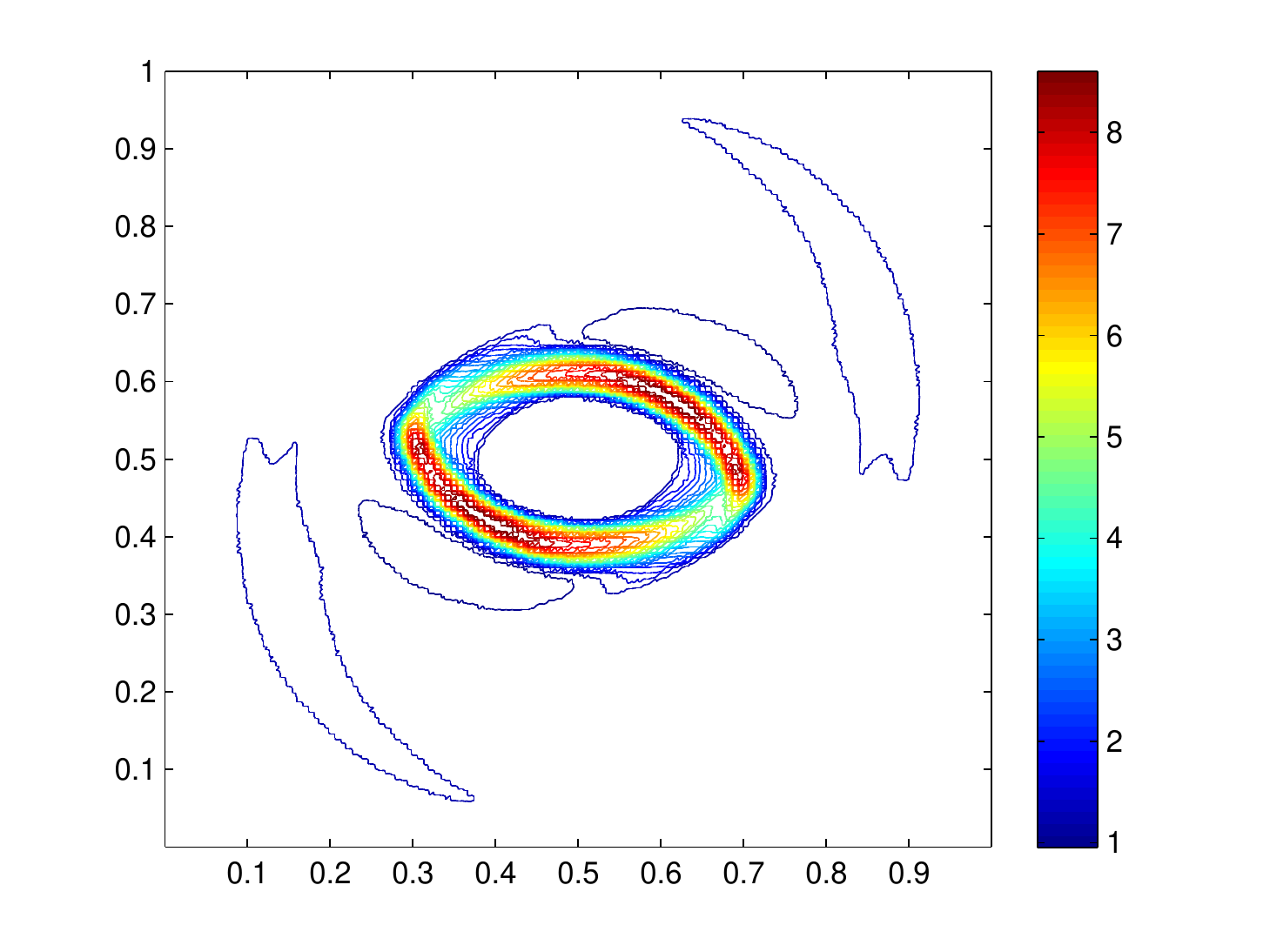,height=2.4in}}
\subfigure[]{\label{fig:press_rotor}\epsfig{figure=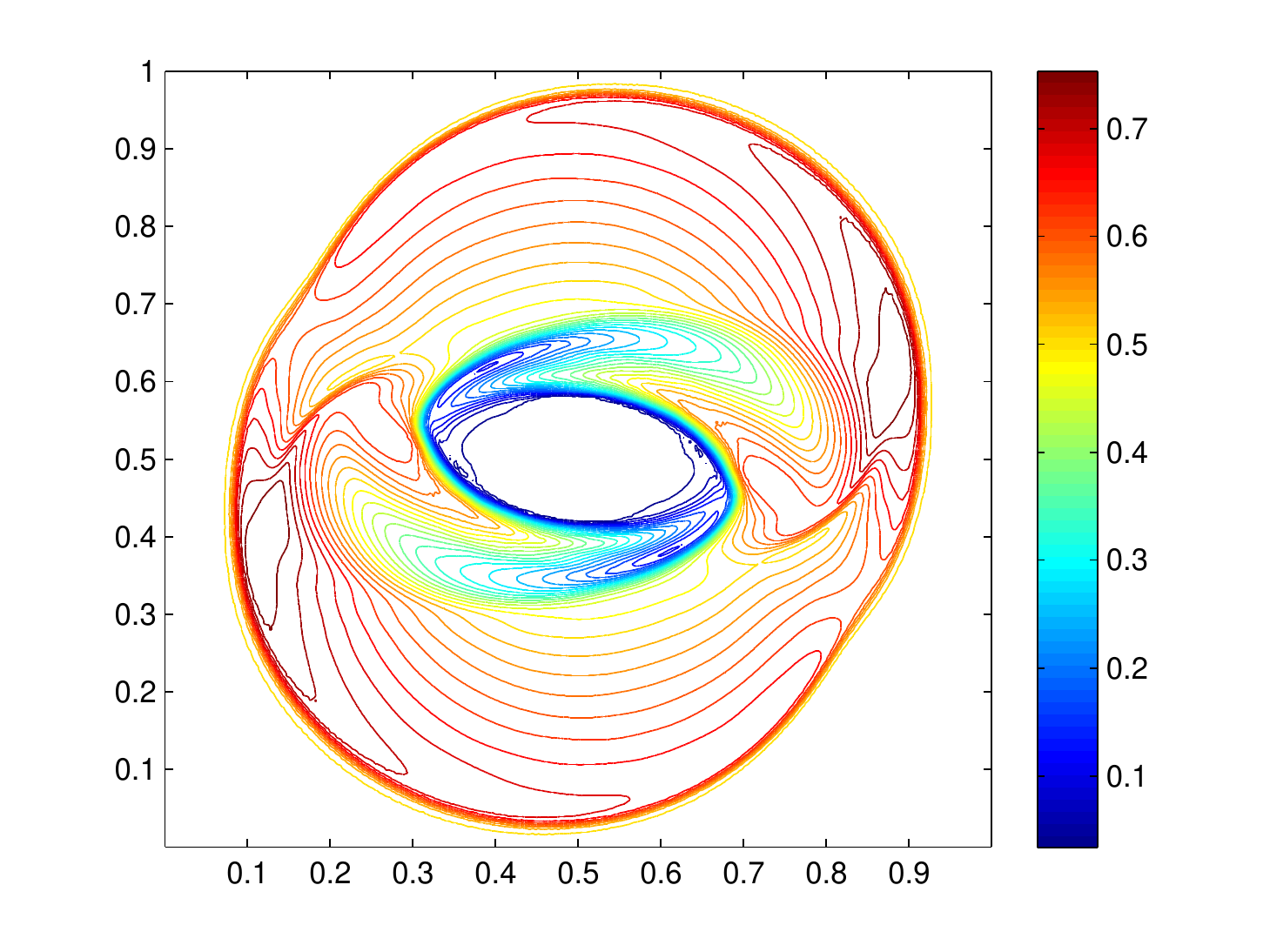,height=2.4in}}
\subfigure[]{\label{fig:B2_rotor}\epsfig{figure=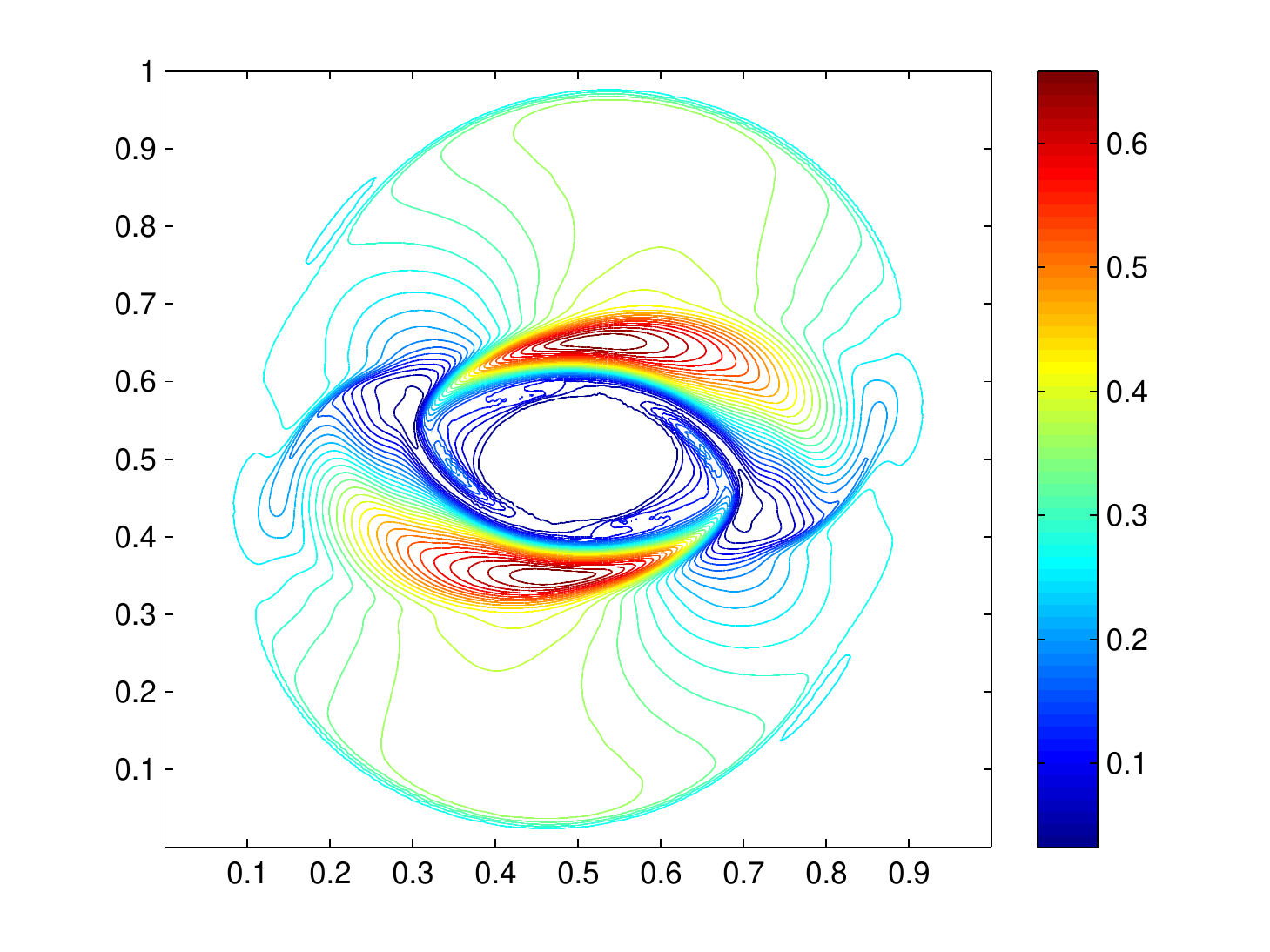,height=2.4in}}
\subfigure[]{\label{fig:mach_rotor}\epsfig{figure=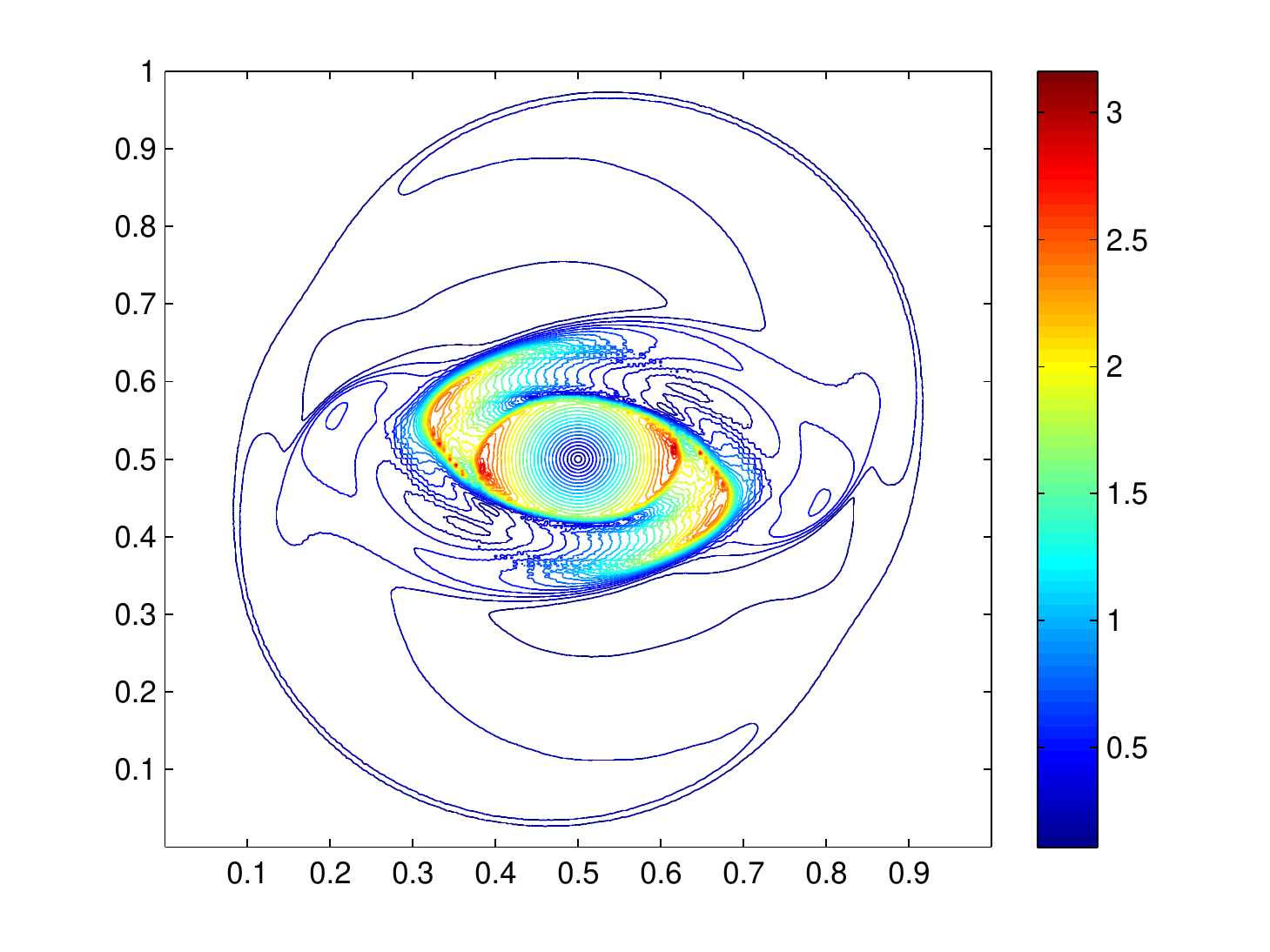,height=2.4in}}
\caption{$P^2$ solution of the rotor problem at time $ t = 0.295$. Thirty equally spaced contours are shown in each plot.
(a) Density $\rho$; (b) Pressure $p_{gas}$; (c) magnetic pressure $(B_x^2 + B_y^2)/2$; (d) Mach number.
}
\label{fig:rotor}
\end{figure}

The solution at time $t=0.295$ is computed. The typical edge length of triangles used to partition the domain is about $\frac{1}{150}$. A CFL number 0.4 is used for calculation. Figure \ref{fig:rotor} plots the numerical result of the density $\rho$, pressure $p_{gas}$, magnetic pressure $(B_x^2 + B_y^2)/2$ and Mach number.
We see that there is virtually no diffusion of the loop's boundaries and no oscillations in the
magnetic pressure within the loop's interior.
The pressure is positive throughout the computational domain.
The degradation in the density variable that was previously reported in \cite{LonDel04} is not seen in our simulation.

\subsection{Blast wave problem}
This test problem is taken from \cite{BalSpi99b}. It was about a spherical strong fast magneto-sonic shock propagates through a low-$\beta$ ($\beta$ = 0.000251) ambient plasma. We use it to show the advantages of the divergence-free reconstruction. The setup of the problem is as follows: on a computational domain $[0, 1] \times [0, 1]$, $\rho =1$, $\vel = 0$, $B_x = 100/\sqrt{4\pi}$, $B_y = B_z = 0$, $p_{gas} = 1000$
within a circle centered at $(0.5, 0.5)$ of radius $R = 0.1$ and  $p_{gas} = 0.1$ elsewhere.
The final simulation time $t = 0.01$.
The typical edge length of triangles used to partition the domain is about $0.0075$. This is a stringent test problem \cite{BalSpi99b}. The pressure is several orders of magnitude smaller than the magnetic energy. A small discretization error in the total energy can produce negative pressure
 near the shock front, as observed by others \cite{Li08,LiXu11}. We used the
negative pressure fix \emph{Strategy 1} in \cite{BalSpi99} to treat this. Briefly, in addition to evolve conservative variables in Eq. $\ref{eq:2d_consv}$, we also update the entropy density on each cell in every numerical time step. If after reconstructing the magnetic field on a cell, the pressure computed from cell average values becomes negative, we derive the updated pressure from entropy and use that to form a new total energy density
which corresponds to a positive pressure. We next use the new total energy density to reconstruct a polynomial approximation to the energy function; while density and momentum are reconstructed by using the average values computed by the base finite volume scheme respectively. We note that this treatment violates conservation of total energy locally. However, we only violate conservation in local regions by an amount that is smaller than the discretization accuracy. And we obtain a numerically consistent and positive pressure which is important for the physics of the problem.

\begin{figure}[!ht]
\centering
\subfigure[]{\label{fig:den_blast}\epsfig{figure=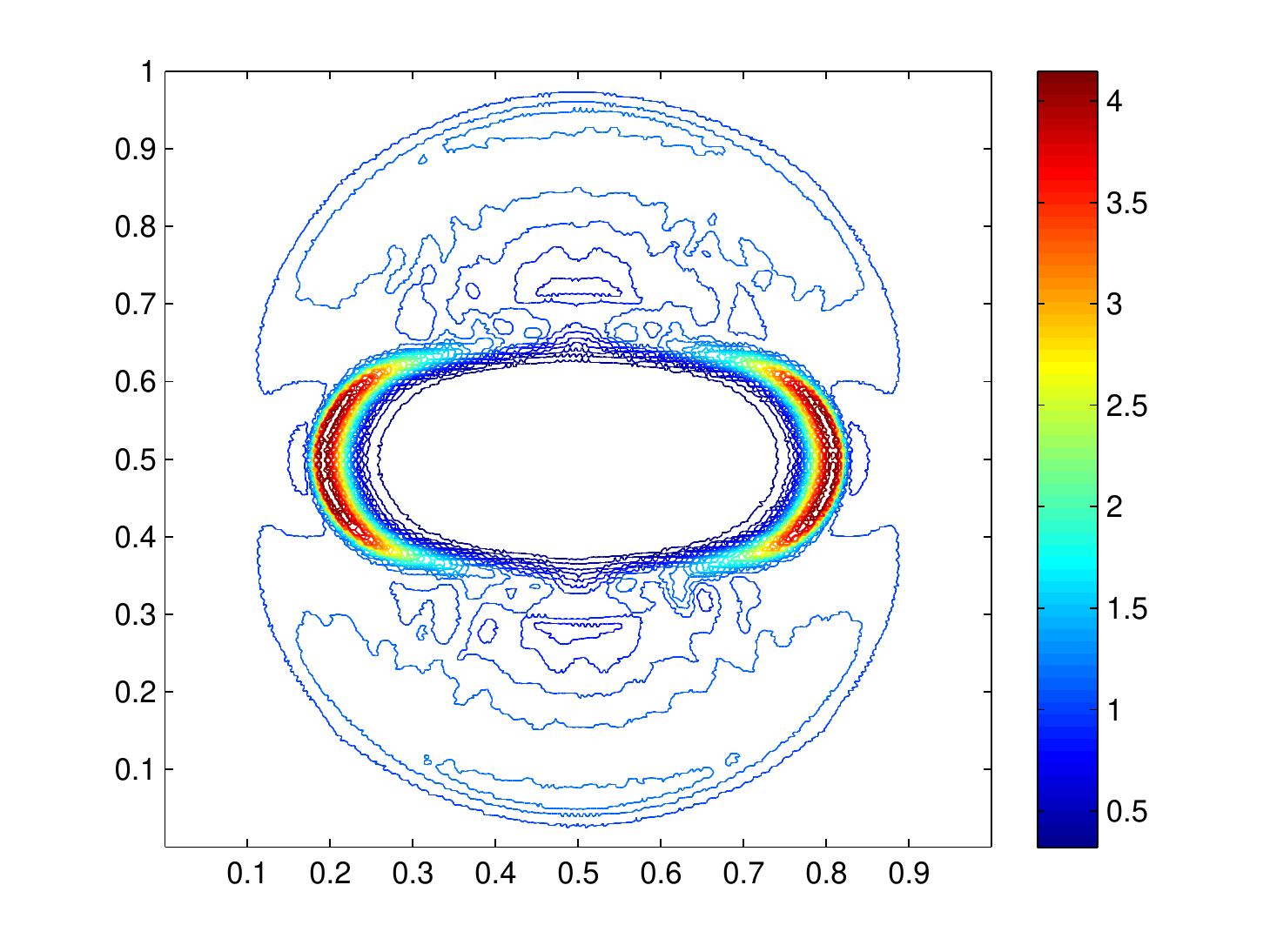,height=2.4in}}
\subfigure[]{\label{fig:press_blast}\epsfig{figure=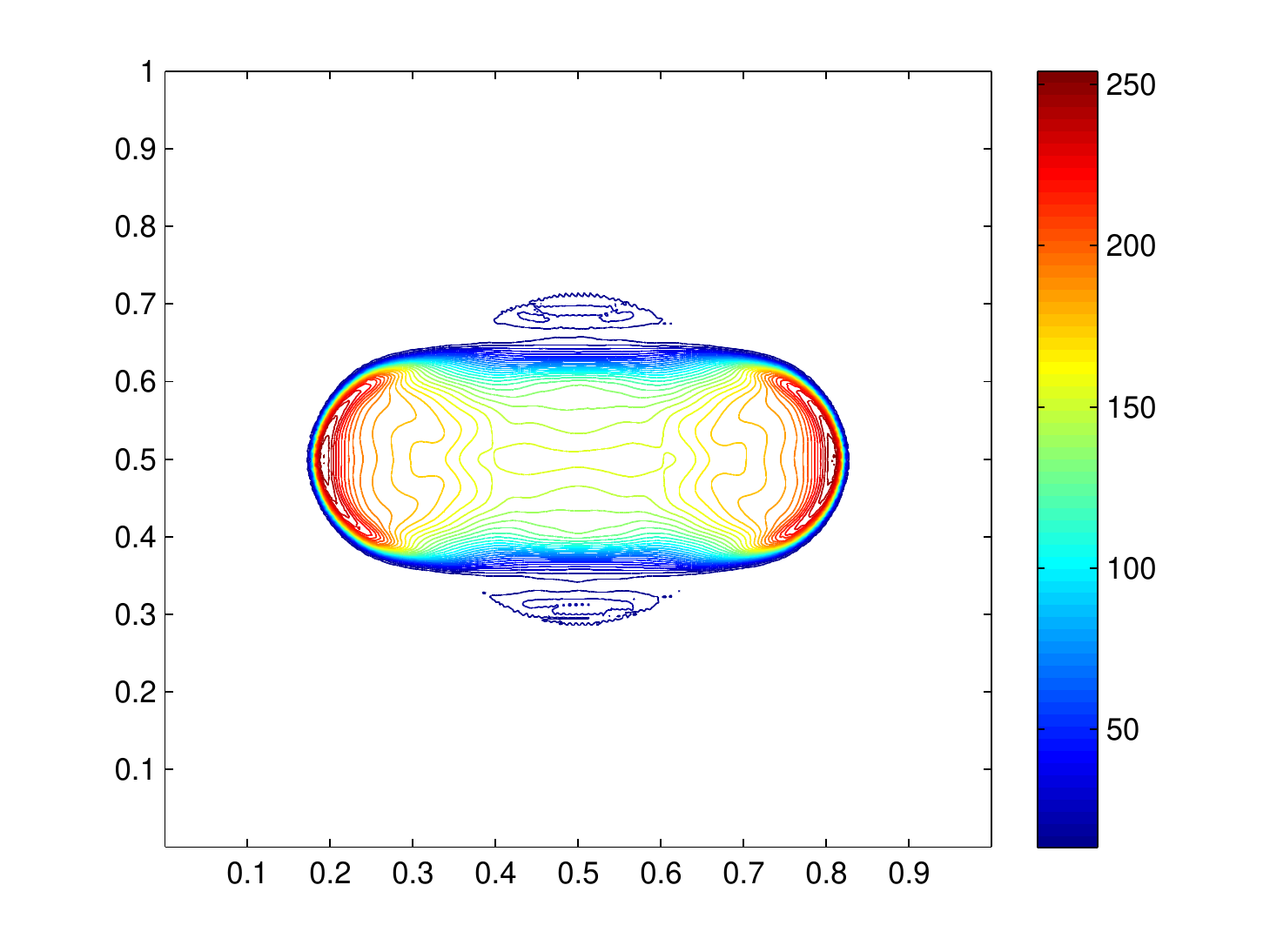,height=2.4in}}
\subfigure[]{\label{fig:B2_blast}\epsfig{figure=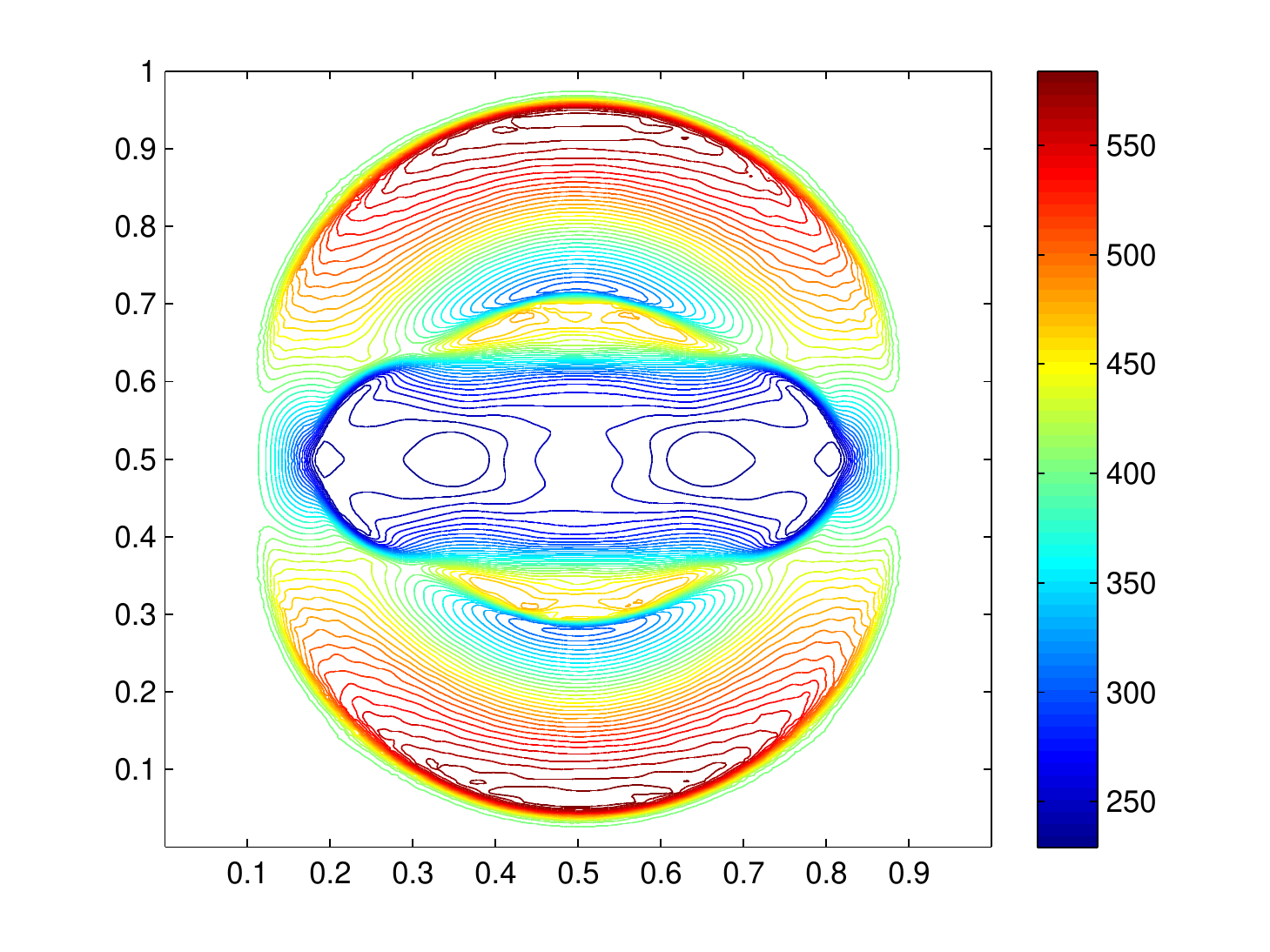,height=2.4in}}
\subfigure[]{\label{fig:mach_blast}\epsfig{figure=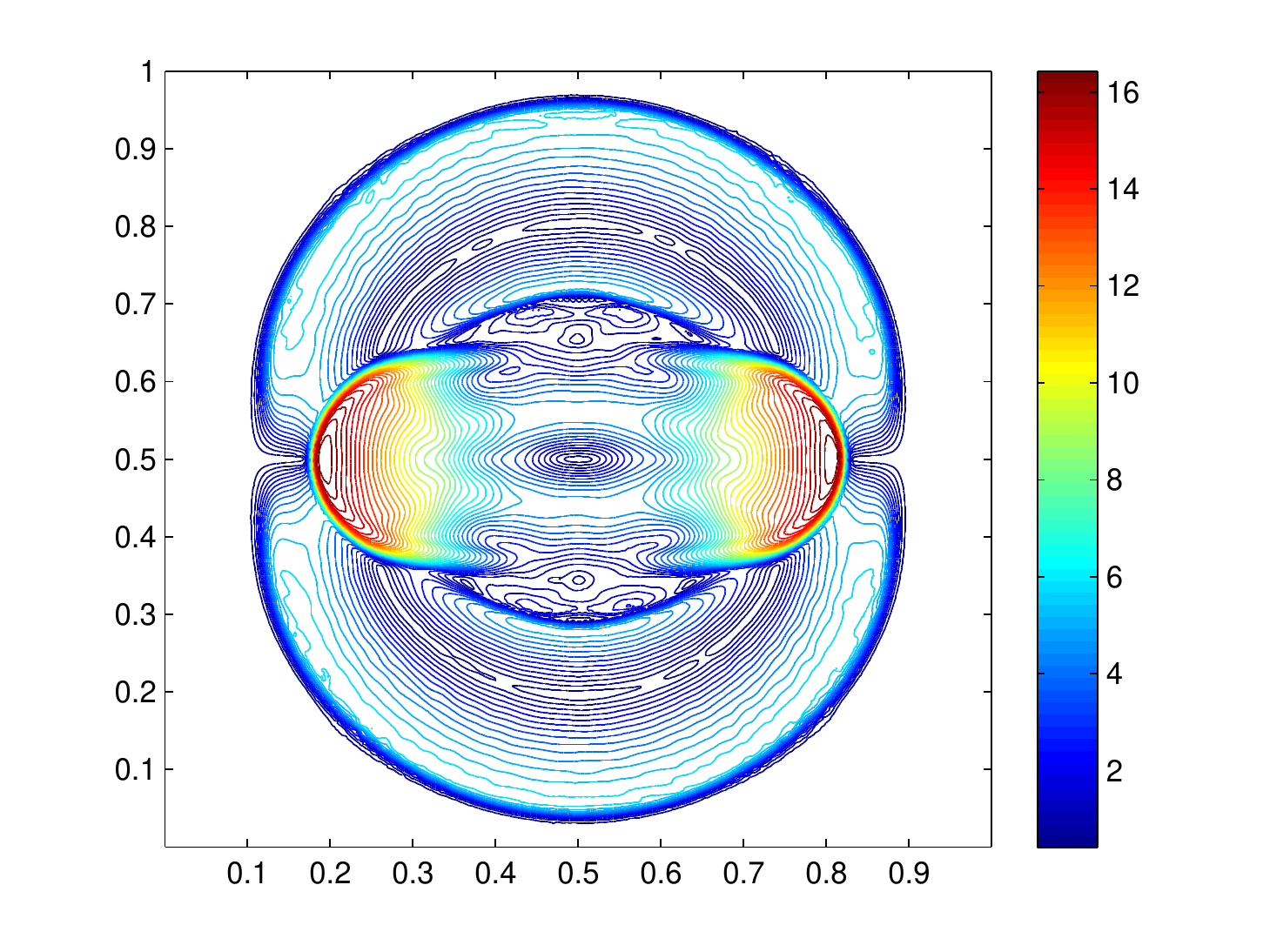,height=2.4in}}
\caption{$P^2$ solution of the blast wave problem at time $ t = 0.01$. Forty equally spaced contours are shown in each plot.
(a) Density $\rho$; (b) Pressure $p_{gas}$; (c) magnetic pressure $(B_x^2 + B_y^2)/2$; (d)
Magnitude of the velocity $\sqrt{u_x^2 + u_y^2}$.
}
\label{fig:blast}
\end{figure}

Figure \ref{fig:blast} plots the numerical result of the density $\rho$, pressure $p_{gas}$, magnetic pressure $(B_x^2 + B_y^2)/2$ and magnitude of the velocity $\sqrt{u_x^2 + u_y^2}$. Owing to the large pressure placed at the center of the domain at the start of calculation, a strong blast wave propagates  outwards, leaving a low density region in the center of the computational domain. We see that there is only minor oscillations in the
density plot. Other fields are resolved nicely.

\subsection{Orsag-Tang problem}
\begin{figure}[!ht]
\centering
\subfigure[]{\label{fig:den_OT05}\epsfig{figure=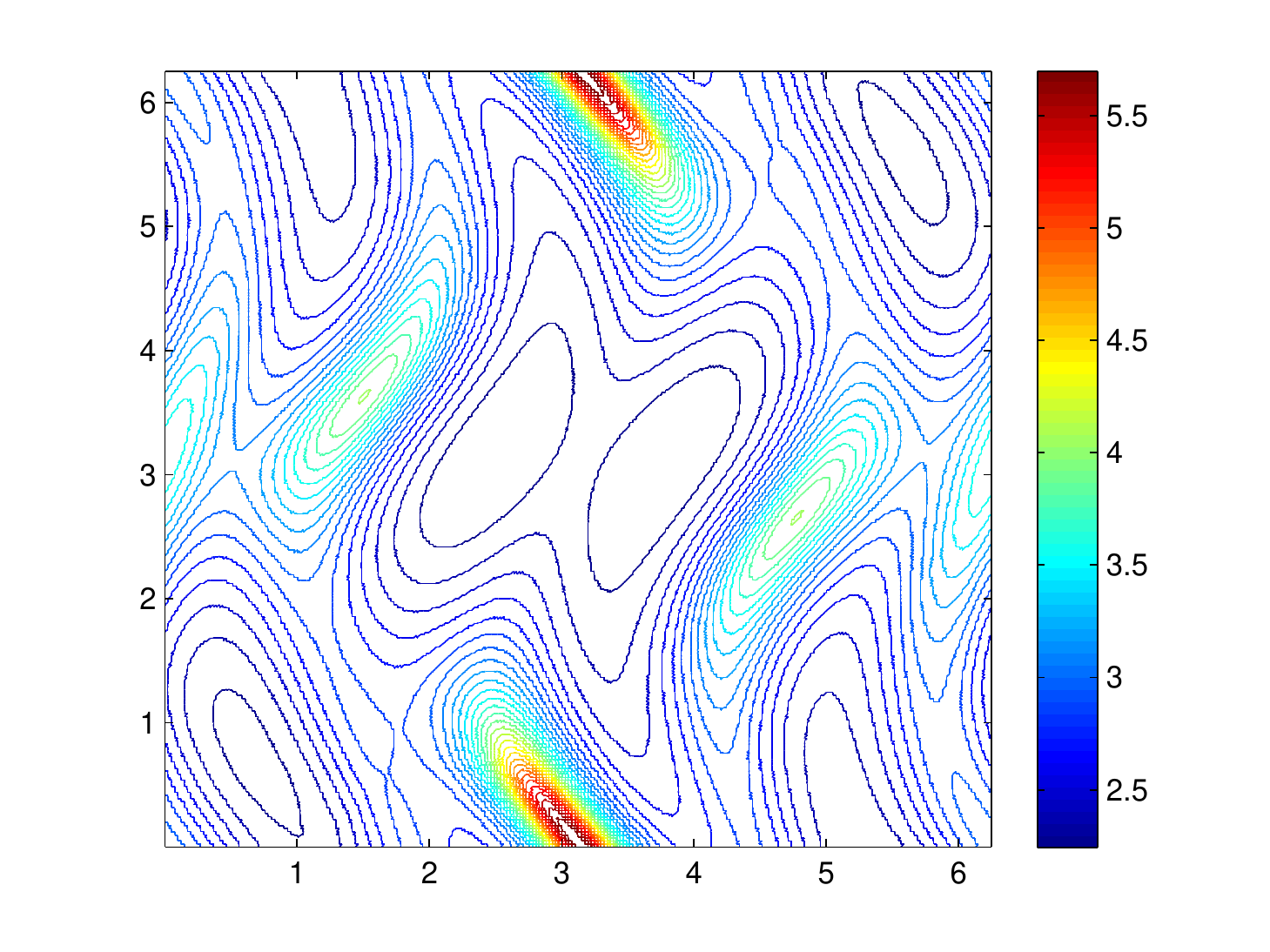,height=2.4in}}
\subfigure[]{\label{fig:den_OT1}\epsfig{figure=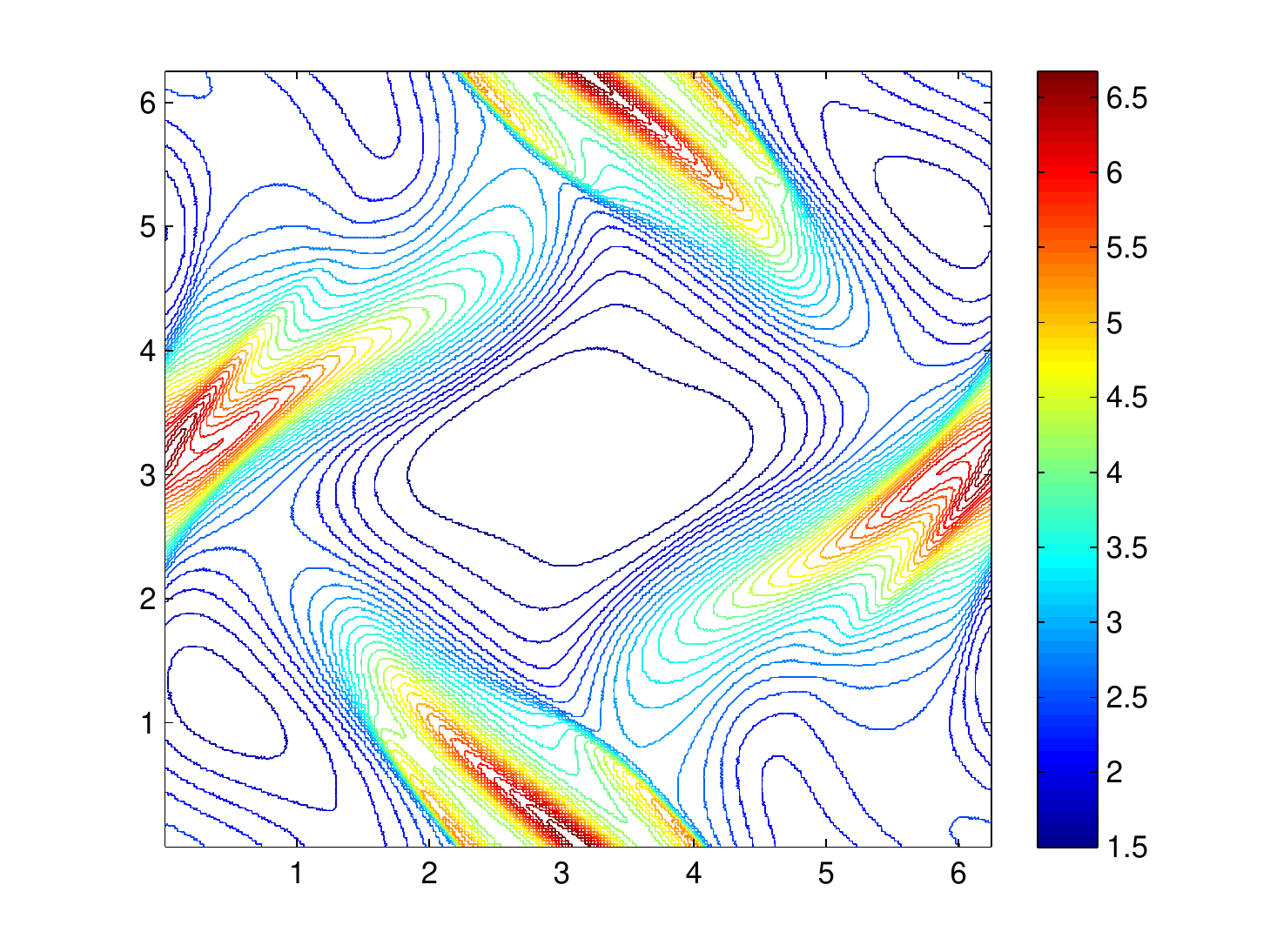,height=2.4in}}
\subfigure[]{\label{fig:den_OT2}\epsfig{figure=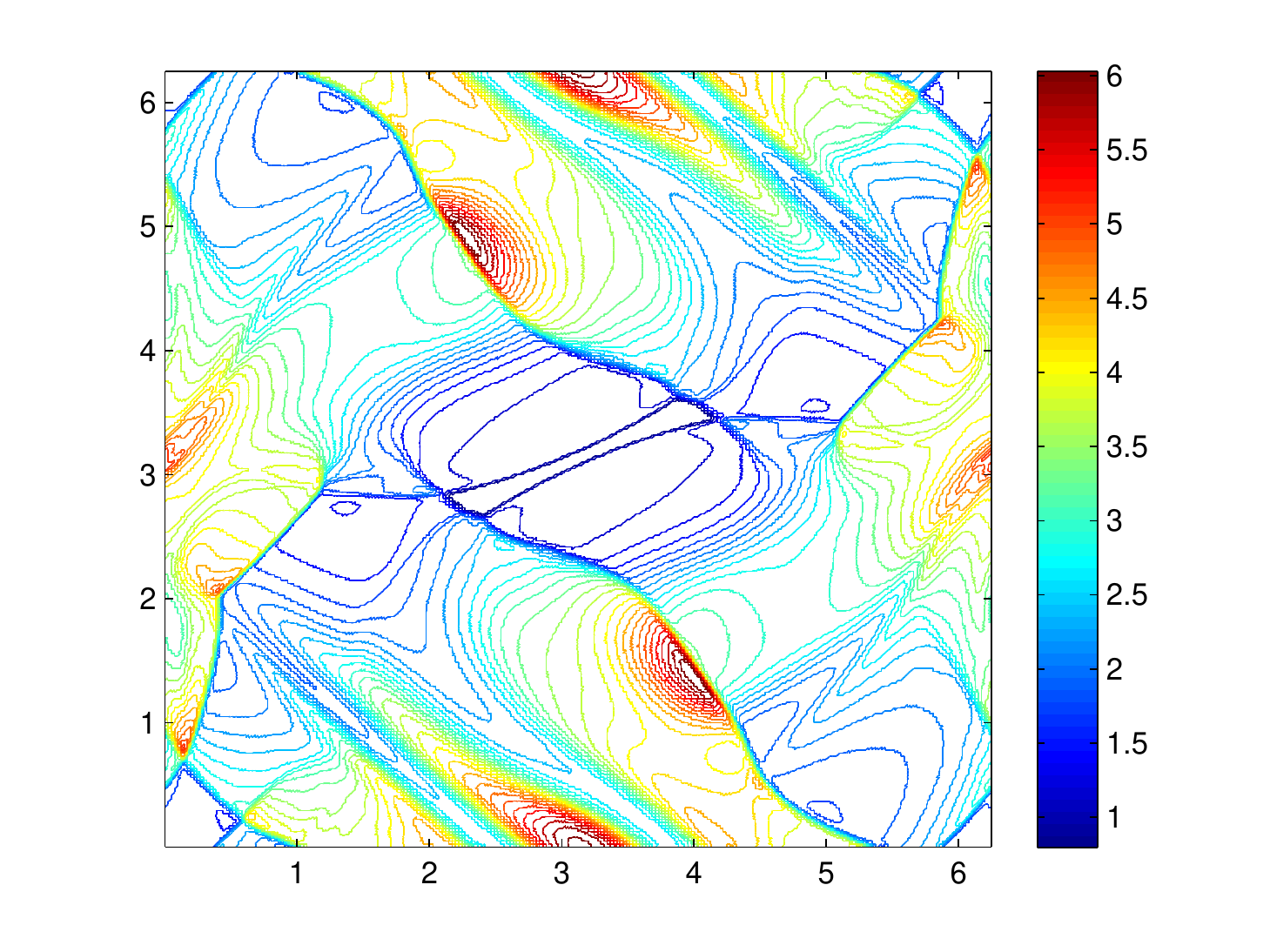,height=2.4in}}
\subfigure[]{\label{fig:den_OTpi}\epsfig{figure=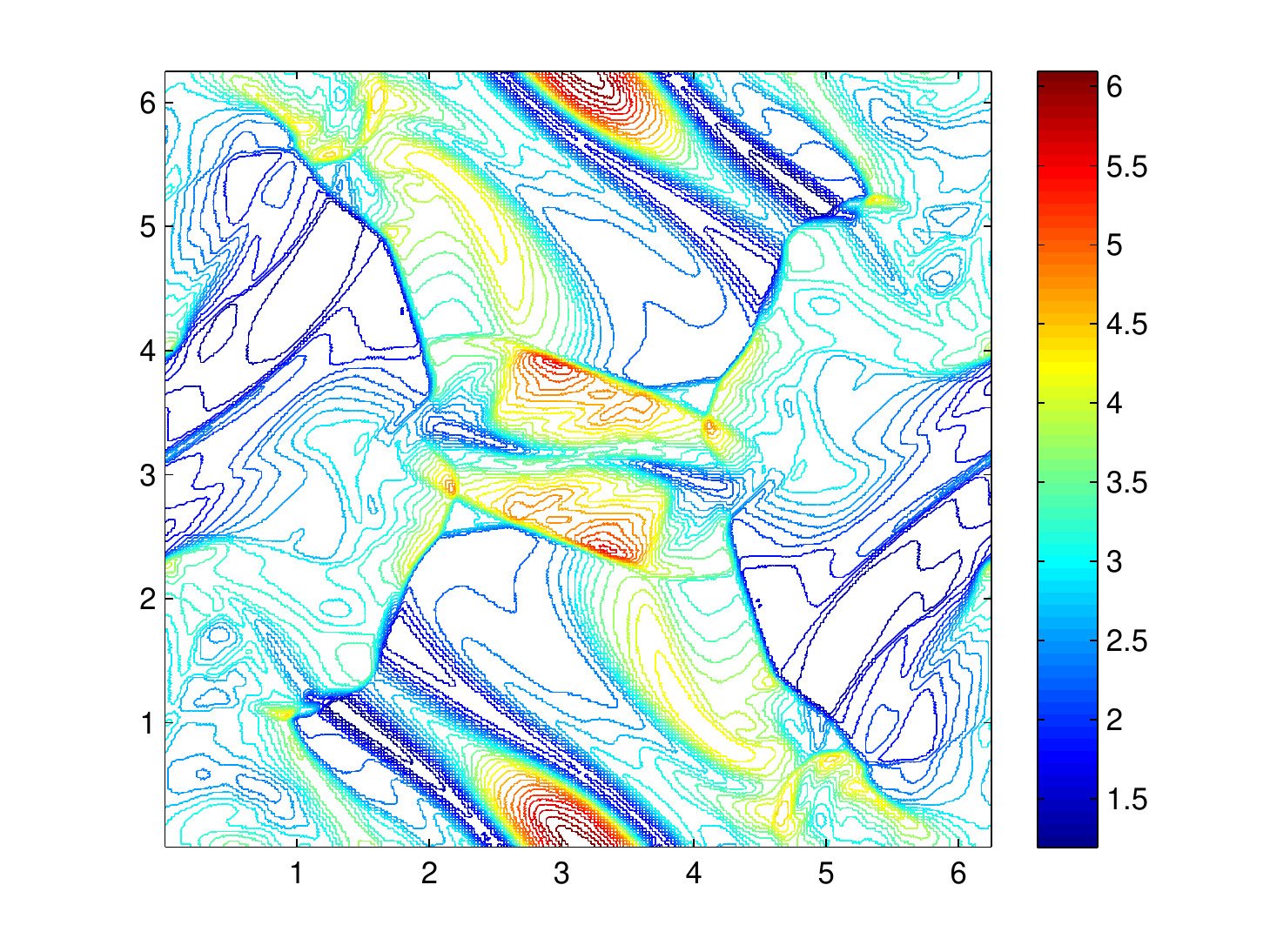,height=2.4in}}
\caption{Orsag-Tang problem. Evolution of $\rho$ over time. Top left: $t = 0.5$;
top right: $t = 1.0$; bottom left: $t = 2.0$; bottom right: $t = 3.14$. 15 equally spaced contours are used.}
\label{fig:orsag_Tang}
\end{figure}

Here we simulate the Orszag-Tang vortex problem \cite{OrzTan79}.
The initial conditions are $u_x = -\sin(y)$ $u_y = \sin(x)$,  $Bx = -\sin(y)$, $By =\sin(2x)$,
 $\rho = \gamma^2$, $p_{gas} = \gamma$, $u_z  = B_z  = 0$.
The computational domain is a square $[0, 2\pi] \times [0, 2\pi]$ with periodic boundary conditions along both boundaries. $\gamma = 5/3$.
The final output time $t = \pi$. The typical edge length of triangles used to partition the domain is about $\frac{1}{256}$.
Starting from a smooth initial condition,  the flow becomes
very complex as expected from a transition towards turbulence gradually.
Figure \ref{fig:orsag_Tang} shows the development of density $\rho$ in the Orszag-Tang vortex problem. Also we report that the density and pressure have remained positive.
No positivity fix was needed for this problem.

\section{Concluding Remarks}
\label{sec:conc}
In this paper we introduced a divergence-free WENO reconstruction-based finite volume method for solving the ideal MHD equations on two-dimensional triangular grids. The proposed method is based on the CT framework and achieves exactly divergence-free magnetic field. Numerical tests show that the proposed schemes have achieved the desired order of accuracy and the third order accurate scheme has been shown to perform very well for shock wave problems. While this paper only implements the second order accurate and the third order accurate schemes, the proposed method in principle can be generalized to three dimensions and to general meshes.

\bibliographystyle{plain}

\end{document}